\documentclass[useAMS,usenatbib]{mn2e}

\usepackage{graphicx,times}
\usepackage{colortbl}
\usepackage{soul,color}
\usepackage[normalem]{ulem}

\def\aap{{A\&A}}		
 
\def\apj{{ApJ}}
\def\apjl{{ApJL}}		
		
\def\aj{{AJ}}
\def\apss{{Ap\&SS}} 
\def\mnras{{MNRAS}}
\def\jpcs{{JPCS}}

\def\nature{{Nature}}

\def\ltorder{\mathrel{\raise.3ex\hbox{$<$}\mkern-14mu
             \lower0.6ex\hbox{$\sim$}}}

\def\r1{$r_1$}

\title[MOCCA Code for Star Cluster Simulations - {}{{IV}}. A new Scenario for IMBH Formation]{MOCCA Code for Star Cluster Simulations - {}{{IV}}. A new Scenario for Intermediate Mass Black Hole Formation in Globular Clusters}
\author[M. Giersz,  N. Leigh, A. Hypki, N. L{\"u}tzgendorf \& A. Askar]{Mirek Giersz$^{1}$\thanks{E-mail: mig@camk.edu.pl (MG)}, Nathan Leigh$^{2,3}$, Arkadiusz Hypki$^{1,4}$, Nora L{\"u}tzgendorf$^5$ and \newauthor  Abbas Askar$^{1}$\\
$^{1}$Nicolaus Copernicus Astronomical Centre, Polish Academy of Sciences, 
ul. Bartycka 18, 00-716 Warsaw, Poland\\
$^{2}$Department of Astrophysics, American Museum of Natural History, Central Park West and 79th Street, New York, NY 10024\\
$^{3}$Department of Physics, University of Alberta, CCIS 4-183, Edmonton, AB T6G 2E1, Canada\\
$^{4}$Leiden Observatory, Leiden University, P.O. Box 9513, 2300 RA Leiden, The
Netherlands\\
$^5$ESA, Space Science Department, Keplerlaan 1, NL-2200 AG Noordwijk, The Netherlands}

\begin{document}

\date{Accepted \ldots. Received \ldots; in original form \ldots}

\pagerange{\pageref{firstpage}--\pageref{lastpage}} \pubyear{2002}

\maketitle

\label{firstpage}

\begin{abstract} 
We discuss a new scenario for the formation of intermediate mass black holes in dense star clusters. In this scenario, intermediate mass black holes are formed as a result of dynamical interactions of hard binaries containing a stellar mass black hole, with other stars and binaries. We discuss the necessary conditions to initiate the process of intermediate mass black hole formation and the influence of an intermediate mass black hole on the host global globular cluster properties. We discuss two scenarios for
intermediate mass black hole formation. The SLOW and FAST scenarios. They occur later or earlier in the cluster evolution and require smaller or extremely large central densities, respectively.  In our simulations, the formation of intermediate mass black holes is highly stochastic. In general, higher formation probabilities follow from larger cluster concentrations (i.e. central densities). We further discuss possible observational signatures of the presence of intermediate mass black holes in globular
clusters that follow from our simulations. These include the spatial and kinematic structure of the host cluster, possible radio, X-ray and gravitational wave emissions due to dynamical collisions or mass-transfer and the creation of hypervelocity main sequence escapers during strong dynamical interactions between binaries and an intermediate mass black hole. All simulations discussed in this paper were performed with the MOCCA Monte Carlo code. MOCCA accurately follows most of the important physical
processes that occur during the dynamical evolution of star clusters but, as with other dynamical codes, it approximates the dissipative processes connected with stellar collisions and binary mergers.   
\end{abstract}

\begin{keywords}globular clusters: general - stars: black holes - methods: numerical
\end{keywords}

\section{Introduction}\label{sec:int}

Recent high resolution observations of globular clusters (GCs) have provided a detailed picture of their physical status, revealing complex phenomena connected with multiple stellar populations, binary evolution and the Galactic tidal field. Despite such great observational progress, many theoretical uncertainties remain in understanding GC formation.  These include the origins of their multiple stellar populations, the properties of their primordial binary populations and the possible presence of both stellar mass and intermediate mass black holes (IMBHs).

IMBHs provide a missing link between stellar-mass black holes (BH) and supermassive BHs (SMBH). There are several definitions of IMBHs available in the literature based on their characteristic mass range but, for the purpose of this paper, only IMBHs with masses between about $100 - 10^5 M_{\odot}$ are considered. Observational confirmation of the existence of IMBHs would have an important impact on a number of open astrophysical problems related to the formation of SMBHs and their host galaxies, the origins of ultraluminous X-ray sources in nearby galaxies and the detection of gravitational waves. In the Milky Way galaxy, the presence of IMBHs in the cores of some GCs has been debated for a 
long time. There are many theoretical and observational arguments in favour of the formation of IMBHs in the centres of Galactic GCs \citep[e.g.][and reference therein]{Lutzgendorfetal2013}, but no \textit{firm} observational confirmation of their presence has been presented to date. Notwithstanding, several claims of their presence have been made, based on applying Jeans' model to detailed photometric and spectroscopic data \citep[e.g.][and references therein]{gebhardt05,vandermarel10}. More concretely, several direct observations of the X-ray emission from distant star clusters have revealed ultraluminous X-ray sources at off-centre positions within their host galaxies \citep[e.g.][]{maccarone08}. These observations suggest the presence of IMBHs with masses of about a few thousand $M_{\odot}$. 

Several widely different theoretical scenarios have been proposed in the literature for the formation of IMBHs in GCs: (1) the formation of very massive Population III stars and their subsequent collapse into a massive BH \citep{MadauRees2001}, (2) the runaway merging of main sequence stars in young and very dense star clusters \citep{Portegiesetal2004, Gurkan2004}, (3) the accretion of residual gas onto stellar mass BHs formed from first generation stars \citep{Leighetal2013a}. A strong motivation for searching for IMBHs is the observed relation between the black hole (BH) mass and the velocity dispersion of its host galaxy \citep[e.g.][and references therein]{FerrareseMerritt2000,Gultekin2009}. Extrapolating this relation to the velocity dispersions characteristic of Galactic GCs, one expects to find central BHs with masses similar to those characteristic of IMBHs.

Scenario (1) invokes the direct collapse of Population III stars formed out of unmagnetized metal-free gas, and implies that the initial mass function (IMF) was extremely top heavy. Stars with masses larger than $200 - 300 M_{\odot}$ might have formed in so-called minihalos with masses about $10^5 M_{\odot}$ at redshifts z about $20$. After only a few Myr, such extremely massive stars would have formed BHs with masses about $200M_{\odot}$, loosing only a small fraction of their mass in the process \citep{FryerWH2001,WhalenFryer2012}.  These BHs could then have acted as the seeds for the formation of IMBHs in GCs. But, doubts remain regarding the validity of this scenario.  For instance, how could such a massive BH formed from a Population III star end up in a GC consisting of Population II stars? Recently, \citet{Belczynskietal2014} discussed extensively the possibility of formation of very massive stars (with masses greater than $150 M_{\odot}$) in star-forming regions. They argue that such massive stars are currently forming, and that they could be IMBH seeds.   

Scenario (2) invokes the runaway merging of massive main sequence (MS) stars. In order for this to occur, the time-scale for mass segregation of the most massive stars (about 100 $M_{\odot}$) has to be shorter than the stellar evolution time-scale for those stars. Otherwise, they will lose a substantial amount of mass before segregating to the cluster centre, where further mass-loss powers cluster expansion. Then, the core will be of low density and consist of relatively low-mass evolved stars, instead of being dense and populated by massive stars. Additionally, the velocity dispersion in a collapsed cluster cannot be larger than a few hundred km/s, otherwise collisions between massive MS stars will be disruptive and result in significant mass-loss.  Regardless, if the right conditions are met, a very massive MS star can form from direct physical collisions between many lower mass MS stars. The resulting collision product will shortly end its evolution as a very massive BH, an IMBH seed.  

Recently \citet{Leighetal2013a} proposed a new model for gas expulsion in very massive star clusters, which might relate to the formation of multiple stellar populations in GCs. The basic idea is connected with gas accretion onto BHs, formed from the most massive stars. The interaction between the BHs and the interstellar medium should speedup and even enhance their mass segregation \citep{Leighetal2014b}, possibly increasing their masses in the process. This could lead to the rapid formation of a dense and (possibly) massive subsystem of BHs. Further evolution can lead to the formation of an IMBH, and/or significantly deplete GCs of their primordial gas reservoirs (via direct accretion onto BHs and/or mechanical/radiative feedback).  

BHs cannot be observed directly. Indirectly, however, they are observable via radio, X-ray and gravitational radiation due to various types of accretion processes. Observations of the radio and X-ray emissions from distant galaxies have led to detections of ultraluminous X-ray sources at off-centre positions, which possibly are young star clusters. \citep[e.g.][and references therein]{Matsumotoetal2001,Soriaetal2010}. These observations are indicative of the presence of IMBHs with masses of about a few thousand $M_{\odot}$. These masses are in agreement with what is expected from extending the observed M - $\sigma$ correlation for supermassive BHs \citep{Magorianetal1998} to lower galaxy masses. 

IMBHs may also be indirectly observable via subtle changes in the cluster structure in their vicinity.  Theoretical investigations and numerical simulations point out that if an IMBH is present in the centre of a GC the velocity dispersion profile (VDP) should rise in the centre, and the central surface brightness profile (SBP) should be shallow \citep[e.g.][and references therein]{BaumgardtMakinoHut2005,TrentiVesperiniPasquato2010,NoyolaBaumgard2011}. Thus, when hunting for IMBHs in GCs, the best candidates are massive GCs with shallow central SBPs and centrally rising VDPs. However, this straightforward prediction should be applied with caution. As was shown by \citet{BaumgardtMakinoHut2005}, the massive evolved stellar population in the core of a GC can also show a rising VDP. Conversely, the presence of a massive stellar mass BH binary in the core can produce a shallow central SBP \citep{Hurley2007}. 

In Section \ref{sec:met}, we briefly describe the Monte Carlo code, called MOCCA, used in this paper to perform simulations of GC evolution. In Section \ref{sec:con}, we describe the initial conditions for our GC models. Next (Section \ref{sec:res}) we describe a new scenario for IMBH formation, and discuss the conditions needed for this to occur. The observational implications of our scenario for IMBH formation in GCs are discussed in Section \ref{sec:obs}. In the final two Sections, we summarise our results and discuss some limitations and future developments of the MOCCA code.

\section{Method}\label{sec:met}

The MOCCA (MOnte Carlo Cluster simulAtor) code used for the star cluster simulations presented here is the Monte Carlo code based on H\'{e}non's implementation of the Monte Carlo method \citep{Henon1971}, which was further substantially developed by Stod\'{o}{\l}kiewicz in the early eighties \citep{Stodolkiewicz1986}. This method can be regarded as a statistical way of solving the Fokker-Planck equation. The basic assumptions behind the Monte Carlo method are: (1) spherical symmetry, which makes it easy to quickly compute the gravitational potential and stellar orbits at any place in the system. This is a very severe assumption, e.g. the evolution of a system with rotation cannot be investigated. In specified spherically symmetric potentials every star is characterized by its mass, energy and angular momentum. Each star moves on a rosette orbit. There is no need for integration of an orbit in the Monte Carlo method, since it is easy to calculate the position of each star along its orbit. In this way, the Monte Carlo method is very fast; (2) cluster evolution is driven by two-body relaxation, and the time step at each position in the system is proportional to the local relaxation time. This is the second reason why the Monte Carlo method is so fast. Generally, Monte Carlo codes cannot follow any physical processes with characteristic time scales much shorter than the local relaxation time.

The MOCCA code \citep{Gierszetal2013} is a heterogeneous code, composed of independent modules (e.g. single and binary star evolution, small N-body scattering, etc.) that together model the entire cluster evolution. It is able to follow most of the important physical processes that occur during the dynamical evolution of star clusters. In addition to the inclusion of relaxation, which drives the dynamical evolution of the system, MOCCA includes: (1) synthetic stellar evolution of single (SSE) and binary (BSE) stars using the prescriptions provided by \citet{HurleyPT2000} and \citet{HurleyTP2002} (from now on we will refer to both codes together as BSE), (2) direct integration procedures for small $N$ sub-systems using the Fewbody code \citep{FCZR2004}, and (3) a realistic treatment of escape processes in tidally limited clusters based on \citet{FukushigeHeggie2000}.  Here, the escape of an object from the system is not instantaneous, but delayed in time. 

The MOCCA code has been extensively tested against the results of N-body simulations \citep[][and references therein]{Gierszetal2013,Heggie2014,Longetal2015}. The agreement between these two different types of simulations is excellent.  This includes the global cluster evolution, mass segregation time scales, the treatment of primordial binaries (energy, mass and spatial distributions), and the numbers of retained neutron stars (NS) and BHs.  The key assumption implemented in the MOCCA code that guarantees this agreement is that, throughout the entire cluster, the time scale for significant evolution is always a small fraction (about 0.001 - 0.01) of the local relaxation time, but larger than the local crossing time.  It is not possible to directly compare results from MOCCA and N-body for extremely large mass ratios. The only analogous N-body simulations presented in the literature are for small N and usually do not contain all the necessary physics - e.g. stellar/binary evolution, an artificially truncated IMF (e.g. \citet{Konstantinidisetal2013}). However, the recent paper by \citet{Leighetal2014b} clearly shows that the results for binary star populations evolving in the presence of an IMBH are the same for N-body and MOCCA. This gives indirect evidence and some confidence that the chosen time-step in the MOCCA simulations is appropriate.  Also, to minimise density fluctuations associated with the most massive stars due to their small numbers, a special procedure was introduced to calculate the probability of binary formation using a smoother average density in the vicinity of the binaries \citep{Giersz2001}. The relaxation of very massive stars via long-range encounters with other mostly low-mass stars, which is pivotal to IMBH formation in our simulations, is treated according to the standard two-body prescription given in \citet{Henon1971}.  Since we do not directly calculate the relative force between every pair of particles at each time step, one might naively expect this approach to introduce some error into our results for IMBH formation.  However, as is shown later in the paper (see Section \ref{sec:sbp}), the spatial and kinematic structures of our model star clusters with IMBH formation reasonably reproduce the results of analogous N-body simulations performed by other authors.

We would like to stress that the Fewbody code models \textit{only} direct gravitational interactions between stars and binaries.  The hydrodynamics, gravitational radiation (GR), tidal effects (distortions and dissipation), mass-loss, etc. are not accounted for. Using the sticky-star approximation, we make the general assumption that, if the stellar radii overlap during an encounter, then the hydrodynamics will be sufficiently non-negligible to bind the stars at a sufficiently small separation that they are well within the stellar envelope of the collision product, which then subsequently settles back on to a standard stellar evolution track on a Kelvin-Helmholtz timescale (which should typically be longer than the total encounter duration, but shorter than the time until a subsequent encounter occurs), with the original components ultimately merging inside the envelope (rapidly - i.e. on a timescale much shorter than the encounter time) to form the core of the new star.  Thus, to first-order, we assume that after a collision occurs, the product is puffy for the remainder of the encounter, but returns to having a radius \textit{normal} for a star of that mass post-encounter (if it avoids a subsequent collision). The subsequent stellar evolution of such a thermally relaxed star is treated by the BSE code.  These assumptions are at least consistent with our current understanding of the relevant physics, and to address them further is beyond the scope of the present analysis.  Notwithstanding, they should eventually be re-visited in more detail.

Importantly, all of the neglected physics of significance to our results act as sources of energy damping during dynamical interactions.  This includes tides and gravitational radiation.  In general, we expect damping to facilitate collisions, and increase the frequency of collisions relative to our calculations. Similarly, in the case of stable or long-lived triples (which we neglect in our simulations), the presence of a distant third companion could reduce the time-scale for a subsequent encounter, increase the collision probability during individual encounters \citep{LeighGeller2012} or even drive the inner binary to shorter periods or even merge due to Kozai-Lidov oscillations \citep{perets09}. Hence, our calculated collision rates should increase if these damping sources and stable triples are included in our analysis. Naively, we expect this to increase the rate of IMBH mass build-up, which will be the subject of future work.

The accuracy and stability of the Fewbody code was extensively tested for very large mass ratios. For the mass ratios of interest to this paper, we have verified that the Fewbody code is able to successfully reproduce the energy and angular momenta exchange rates presented in \citet{Sesanaetal2006}.

The MOCCA code provides as many details as N-body codes, while being much faster. It can follow the time evolution and movement of particular objects, and needs only about a day to complete the evolution of a realistic globular cluster. So, instead of just one N-body model, several hundreds of MOCCA models can be computed (with different initial conditions) in a comparable amount of time. Thus, the MOCCA code is ideal not only for dynamical models of particular clusters, but also large surveys. 

\section{Initial Conditions}\label{sec:con}

The new scenario for IMBH formation presented in this paper appeared as a byproduct from two earlier projects. The aim of the first project \citep{Leighetal2013b} was to explain the observed correlations: between cluster mass and binary fraction - the larger the cluster mass, the smaller the binary fraction \citep{Miloneetal2012}; and mass function (MF) power-law index and cluster concentration - the more concentrated the cluster, the larger the MF power-law index \citep{DeMarchietal2007}. The aim of the second project \citep{Leighetal2015a} was to constrain the initial properties of primordial binaries by comparison with observations \citep{Miloneetal2012} and to check if star cluster simulations with initial conditions drawn from the invariant \citet{Kroupa1995} distributions are able to recover the observed spatial distributions of binaries. All together, for these projects about 400 models of star clusters (SC) were simulated by the MOCCA code. 

\begin{table*}
\begin{minipage}{120mm}
  \caption{Initial conditions for the MOCCA simulations.}
  \label{tab:models}
\begin{tabular}{|l|c|c|c|c|c|c|}
\hline
N & $R_t$ (pc) & $R_t/R_h$ & $f_b$ & $a_{max}$ (AU) & IMF & Kicks (km/s) \\
\hline
$N=5.0x10^4$ & 38 & 20, 25, 30 & 0.1 & 100 & IMF2, IMF3 & 0, 265 \\
King $W_0 = 6$    &  & 50, 75  &  & Period &  & Fallback \\
\hline
$N=1.0x10^5$ & 69 & 10, 20 & 0.1 & 100 & IMF3 & 265 \\
King $W_0 = 6$    &  &  &  & Period & IMF2-m & Fallback \\
\hline
$N=2.0x10^5$ & 69 & 10, 20 & 0.1 & 100 & IMF3 & 265 \\
King $W_0 = 6$    &  &  &  & Period & IMF2-m & Fallback \\
\hline
$N=3.0x10^5$ & 35, 50 & 35, 50, 60 & 0.1, 0.3 & 100, 200 & IMF2, IMF3 & 0, 265 
\\
King $W_0 = 6$    & 69 & 75, 100 & 0.7, 0.95 & 400, Period & IMF2-m & Fallback \\
\hline
$N=1.8x10^6$ & 125 & 60, 75, 90 & 0.1 & 100 & IMF2, IMF3 & 0, 265 \\
King $W_0 = 6$    &  & 100, 125 &  & Period  & IMF2-m & Fallback \\
\hline
$N=5.5x10^5$ & 181, 98, 68 & 50, 100 & 0.95 & Period & IMF2, IMF3 & 265 \\
$N=6.9x10^5$ &       &            &     &     &            & \\
King $W_0 = 6$    &  &   &  &   &    &  \\
\hline
$N=1.1x10^6$ & 220, 120, 75 & 50, 100, 200 & 0.95 & Period & IMF2, IMF3 & 265 \\
King $W_0 = 6$    &  & &  &  &  & \\
\hline
$N=6.6x10^4$ & 3.6 & 6.8 & 0.0, 0.1 & 100 & IMF2  & 190 \\
King $W_0 = 6$    &  &  & 0.95 & Period  &  & Fallback \\
\hline
\end{tabular}
\vskip 0.2cm
 \textit{Notes:}
BH and NS kicks are the same, except the case of mass fallback \citep{Belczynskietal2002}. IMF2 - two segmented IMF \citep{Kroupa2001}, IMF3 - three segmented \citep{Kroupaetal1993}, IMF2-m - modified by changing $\alpha_{low}$ and $\alpha_{high}$, Period - binary period distribution \citep{Kroupa1995}. $f_b$ - binary fraction, $N$ - number of stars, $R_t$ - tidal radius, $R_h$ - half-mass radius, $f_b$ - binary fraction, $a_{max}$ - maximum value for semi-major axis (distribution uniform in $log(a)$).
\end{minipage}
\end{table*}

The parameters of models run for the aforementioned projects are listed in Table \ref{tab:models}. Generally, more massive models have larger concentrations (measured as the ratio between tidal and half-mass radii $C = R_t/R_h$ - i.e. not the standard concentration parameter given in, for example, \citet[][updated 2010]{Harris1996}) with some models having initial concentrations $C$ as large as 125. Roughly half of all models had binary fractions $f_b$ of 0.1, while $f_b$ varied from 0.3 to 0.95 in the other half. The adopted binary period distribution was uniform in the logarithm of the semi-major axis with a cut-off at 100AU for all standard models, and either 200AU or 400AU for the other models. For models with large binary fractions, however, the \citet{Kroupa1995} period distribution was used instead, with a cut-off at $log(P/days)=8.3$. Also, we adopted different IMFs for some of our models: the Kroupa canonical IMF - a two segmented IMF \citep{Kroupa2001}, the Kroupa standard IMF - a three segmented IMF \citep{Kroupaetal1993}, and the two segmented modified Kroupa IMF (with different power-law indices above and below the break mass). The minimum and maximum masses for all IMFs were taken to be $0.08 M_{\odot}$ and $100.0 M_{\odot}$, respectively. A metallicity of $Z=0.001$ was adopted for all models. Supernovae (SN) natal kick velocities for BHs were modified for some models according to the mass fallback procedure described by \citet{Belczynskietal2002}.

\section{Results}\label{sec:res}

While analysing the simulation data for the first project \citep{Leighetal2013b}, it was unexpectedly noticed that a substantial buildup of BH mass occurs in some models (see Fig. \ref{fig:imbh}). We find two different regimes of BH mass growth. The first begins at later times in the cluster evolution (usually close to the core collapse time) and has a small rate of mass accretion (SLOW scenario). The second begins near the very beginning of our simulations (usually just after the formation of NSs and BHs via supernovae explosions and formation of a dense self-gravitating BH sub-system), with a very high rate of the mass buildup (FAST scenario). A detailed d1iscussion of these phenomena is postponed to the next subsection. We begin by concentrating on the dynamical mechanisms that lead to the formation of NSs and stellar mass BHs, which ultimately act as the seeds for IMBH formation in the SLOW scenario.

At the end of this Section, we define the terminology used from now on in the paper.  A \textit{hyperbolic collision} refers to a physical collision between two single stars, assuming the sticky-star approximation (i.e. if the radii of two stars overlap, they are assumed to collide and form a single new star).  A \textit{collision} refers to a physical collision (sticky-star approximation) between two or more stars during dynamical interactions between a binary star and either another single star or binary.  A \textit{merger} refers to binary coalescence driven by stellar or binary evolution. Mergers can be forced by earlier binary-mediated dynamical interactions.  These serve to make binaries harder and/or more eccentric, or drive direct collisions.  Thus, collisions and mergers both involve binaries, but hyperbolic collisions involve only single stars.

The time axis of most of the figures presented in this paper extend beyond a Hubble Time - in some models up to 80 Gyr! Such exceedingly long evolutionary time scales are irrelevant for direct comparison to Galactic GCs. However, they offer important insight regarding what to expect for the spatial structure of dissolving clusters containing an IMBH, which could help to identify them observationally \citep[e.g.][]{BanerjeeKroupa2011}.

\subsection{Neutron Stars and Stellar Mass Black Holes}\label{sec:nsbh}

Generally, it is widely assumed that most NSs and BHs in GCs, if they are present, formed due to ordinary stellar evolution.  This leads to SN explosions very early on in the cluster evolution, which impart significant natal kicks to NSs and BHs, with velocities ranging from a few to several hundreds km/s. NSs and BHs that receive kicks larger than the escape velocity are quickly removed from the system. Others remain in the cluster, and experience dynamical interactions with other stars and binaries that strongly shape the cluster evolution \citep{Contentaetal2015} . It is believed that new NSs and BHs can form via collisions between stars only in extremely dense environments \citep[e.g.][and reference therein]{PapishPerets2015}, where the collision rate is sufficiently large. In this section, we consider the possibility of forming NSs and BHs in less dense environments, not only due to physical collisions during single-single, single-binary and binary-binary encounters, but also due to the mergers of binaries via normal stellar/binary evolution or dynamically-forced channels. 

\begin{figure}
{\includegraphics[height=11cm,angle=270,width=8.5cm]{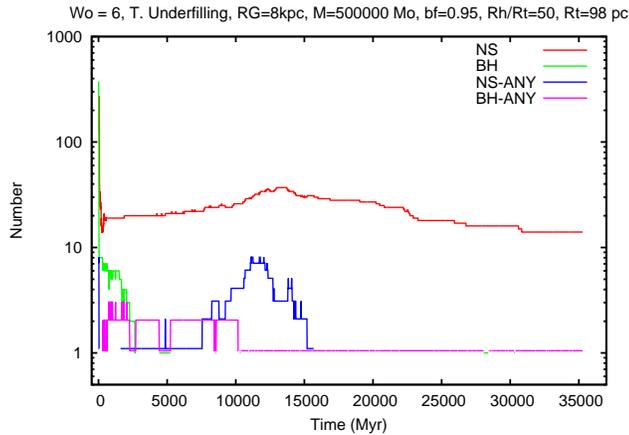}}
    \caption{The numbers of different kinds of objects containing NSs and/or BHs as a function of time, for models in which an IMBH forms. The model parameters are given in the figure title. NS - solid line, BH - long dashed line, NS-ANY - short dashed line, BH-ANY - dotted line. Here, ANY refers to the MS and to all stars that have evolved away from the main sequence.}
\label{fig:n_ns_bh}
  \end{figure}

\begin{figure}
{\includegraphics[height=11cm,angle=270,width=8.5cm]{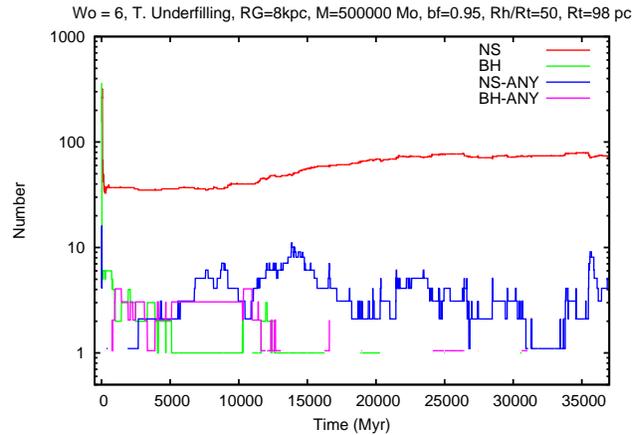}}
    \caption{The numbers of different kinds of objects containing NSs and/or BHs as a function of time, for models in which an IMBH does not form. The model parameters are given in the figure title. The various types of interactions are described in the caption for Figure \ref{fig:n_ns_bh}}. 
\label{fig:no-n_ns_bh}
  \end{figure}

\begin{figure}
{\includegraphics[height=11cm,angle=270,width=8.5cm]{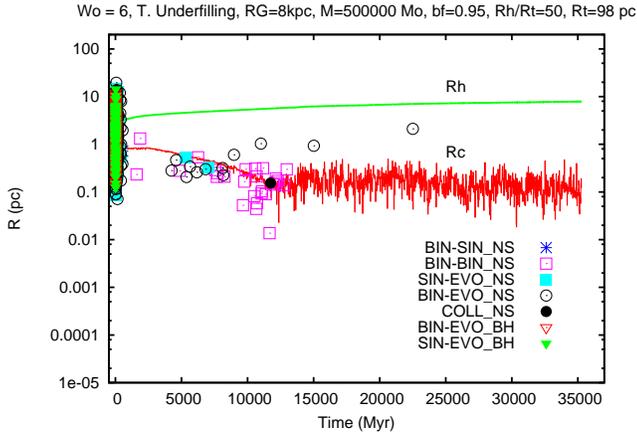}}
    \caption{The position with respect to the centre of the cluster at which NSs and BHs are formed for models with IMBH formation. The model parameters are given in the figure title. $R_c$ and $R_h$ are the core and half-mass radii, respectively. BIN-SIN: binary-single dynamical interaction during which a collision occurs, BIN-BIN: binary - binary  dynamical interaction during which a collision occurs, COLL: single - single collision in hyperbolic encounter, SIN-EVO: single star evolution and BIN-EVO: binary star evolution.}
\label{fig:r_ns_bh}
  \end{figure}

\begin{figure}
{\includegraphics[height=11cm,angle=270,width=8.5cm]{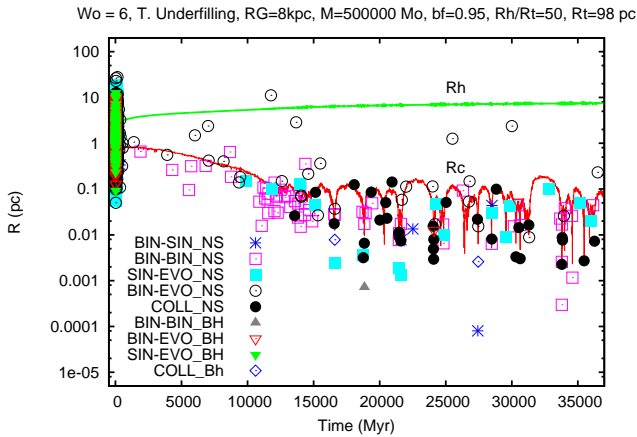}}
    \caption{The position with respect to the centre of the cluster at which NSs and BHs are formed for models without IMBH formation. The model parameters are given in the figure title. $R_c$ and $R_h$ are the core and half-mass radii, respectively. The various types of interaction are described in the caption for Figure \ref{fig:r_ns_bh}.}
\label{fig:no-r_ns_bh}
  \end{figure}

In Figs. \ref{fig:n_ns_bh} and \ref{fig:no-n_ns_bh}, we show the evolution in the numbers of different kinds of objects containing NSs and/or BHs, both for models featuring IMBH formation and those that do not. Evidently, most NSs and BHs are quickly evacuated from the system because of the large natal kicks they receive (the presented models have a natal kick velocity dispersion equal to 265 km/s for both NSs and BHs). Only about 20-30 NSs and a few BHs are left, which then begin to interact dynamically with other stars and binaries. In both figures, the numbers of NSs experiences a sharp initial decrease, followed by a slow rise at later times. This is because new NSs are formed due to collisions during dynamical interactions and mergers due to binary evolution-mediated mass transfer (see Figs. \ref{fig:r_ns_bh} and \ref{fig:no-r_ns_bh}), with the latter dominating substantially over the former. 

When the mass of BH is greater than about 100$M_{\odot}$ and IMBH mass growth begins (see Fig.  \ref{fig:n_ns_bh} - time around of 15 Gyr), all other BHs are removed from the system due to strong dynamical interactions with the IMBH. Only one BH remains in the system, slowly increasing its mass. Similarly, the number of NSs starts to decrease, albeit at a slower rate. If an IMBH does not form, on the other hand, a very different situation arises. In these models, the number of NSs and BHs (both single and in binaries) is much larger at later times. Consequently, a much wider variety of exotic objects containing NSs and BHs are present. This leads to an interesting constraint, which can be verified observationally - that the number of BHs (or, more generally, X-ray binaries) observed in GCs can serve as an indicator of the presence of an IMBH. Systems with an IMBH should not contain any stellar mass BHs. This conclusion is in agreement with the findings of \citet{Leighetal2014b}. It is worth stressing that not all of the NSs and BHs present in our simulations formed via SN explosions near the very beginning of the cluster lifetime. They also form later on in the cluster evolution, either via collisions also hyperbolic collisions) or binary mergers/mass-transfers. 

The radial positions at which NSs and BHs are formed within the cluster during its evolution are show in Figs. \ref{fig:r_ns_bh} and \ref{fig:no-r_ns_bh} for clusters with and without IMBHs, respectively. The BHs are almost exclusively formed at the beginning of the cluster evolution due to stellar and binary evolution of massive MS stars. In contrast, NSs are also formed later on in the cluster evolution, but here they are mainly formed via collisions during binary-binary dynamical interactions or via single and binary star evolution. The NS formation rate is similar in both sets of models (i.e. with and without an IMBH) up to the time of core collapse or IMBH formation. 
The times at which these NS form, which stretch over many Gyr, are much larger than the evolution time for the massive MS stars formed at time T = 0.  Even at these late times, stellar evolution can be rejuvenated by physical collisions, which increase the stellar mass. In more complicated ways, binary evolution can have a similar effect: earlier dynamical interactions of binaries can make a binary harder, or lead to collisions during interactions.  Thus these figures implicitly indicate very clearly the complicated interplay between stellar
dynamics and stellar evolution.
For models in which an IMBH is formed, the late-time formation of NSs and/or BHs is strongly inhibited. The core radius remains large, and the probability of strong dynamical interactions is low. Furthermore, most of the massive objects (stars or binaries) that act as the progenitors for NSs and/or BHs are efficiently removed from the system due to dynamical interactions with the IMBH. A different situation arises in models without IMBH formation. Here, the cluster reaches the gravothermal oscillation phase, such that the core bounces in and out of a very dense state. This environment is ideal for frequent and strong dynamical interactions. Indeed, a substantial number of NSs and a few BHs are typically formed during this phase of cluster evolution, only some of which form via normal stellar/binary evolution. Again, early physical collisions and/or binary dynamical interactions stimulate or induce accelerated binary and/or single star evolution. In the case of binaries this can occur via the hardening of binaries due to dynamical interactions, provided collisions do not happen first. Binary hardening ultimately increases the probability of NS/BH formation due to single/binary evolution. Interestingly, a significant number of NSs formed via binary evolution are found well outside the core, and even beyond the half-mass radius. The progenitor binaries experienced a strong dynamical interaction with the IMBH deep in the cluster core and were subsequently kicked to large clustercentric radii before (or after) NS formation due to single/binary evolution occurred. This suggests an interesting observational constraint for the presence of IMBHs in globular clusters:  Clusters observed to harbour a significant fraction of their NSs beyond $R_h$ could be more likely to host an IMBH.

\begin{figure}
{\includegraphics[height=11cm,angle=270,width=8.5cm]{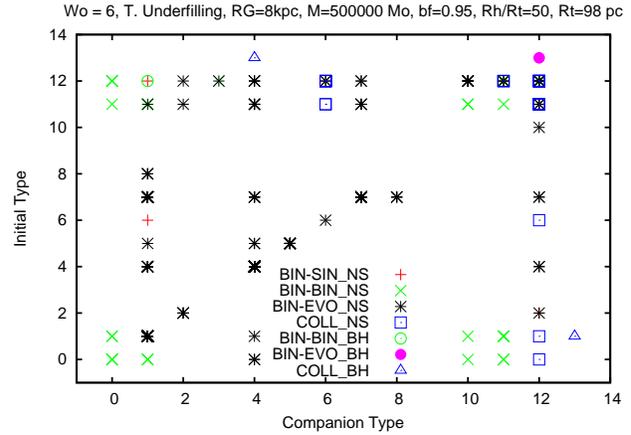}}
    \caption{The different types of interacting stars that will form a NS or BH via a merger/collision for models without IMBH formation. The model parameters are given in the figure title. Initial Type refers to the initial type of a star, which will ultimately merge/collide to form a NS or BH. Companion Type refers to the type of star acting as a binary companion to the colliding star. The different types of interactions are described in the caption for Figure \ref{fig:r_ns_bh}.}
\label{fig:no-t_ns_bh}
  \end{figure}

The conclusions described in the above paragraph regarding the importance for NS/BH formation via collisions, mergers or binary mass transfer (due to both dynamical interactions and normal stellar evolution) are further supported by Fig. \ref{fig:no-t_ns_bh}. 
These figure shows the types of stars which will form NSs or BHs during either stellar/binary evolution or dynamical interactions. The various star types are the same as used in the BSE code \citep{HurleyPT2000}, namely: 0 and 1 -- MS, 2 - 9 -- evolved stars, 10 - 12 -- white dwarfs (WDs), 13 -- NS, 14 -- BH. NSs created in the course of the cluster evolution, long after the initial SN explosions of the most massive stars, are primarily formed due to binary evolution in binaries containing a MS star and an evolved star, containing two evolved stars or containing a WD and an evolved star. There are also merger events connected with gravitational wave radiation (GR) in WD-WD and NS-WD binaries in the case of BH formation. In order for orbit shrinkage due to GR to be efficient, the binary semi-major axis must be on the order of a solar radius. Such small semi-major axes may result either from pure binary evolution involving a common envelope stage (about 30\% of all WD-WD GR mergers), or from binary evolution forced by dynamical interactions with other stars and/or binaries (frequently connected with the exchange of a MS star into a binary and then followed by a common envelope stage). Dynamical interactions can lead to gradual binary hardening and substantial eccentricity pumping, until the semi-major axis is sufficiently small and the eccentricity sufficiently large for GR to become efficient. Usually, these dynamical interactions are not sufficiently strong to make binaries escape from the cluster, but they are strong enough to eject them from the core. Interestingly, there is a substantial number of physical hyperbolic collisions between a WD and an evolved star and even between two WDs. Generally, NSs and BHs that form due to either collisions during dynamical interactions or mergers and binary mass transfer (connected with stellar evolution) have all types of stars as their progenitors.

\subsection{A New Scenario for IMBH Formation}\label{sec:imbh}

\begin{figure}
{\includegraphics[height=11cm,angle=270,width=8.5cm]{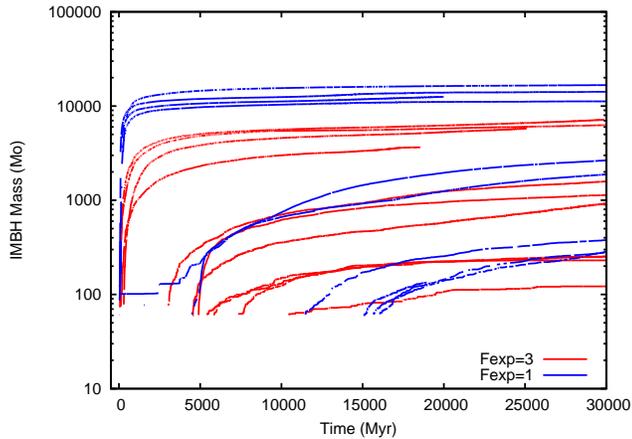}}
    \caption{IMBH mass buildup for models with IMBH formation. Dashed lines - models with reduced mass accretion onto the IMBH (25\% of the standard BSE setup) and reduced (relative to the default) star expansion after merger events ($1/3$ of the standard Fewbody setup). Solid lines - models with the standard prescription for IMBH mass accretion rates and post-merger expansion. See description in the text.}
\label{fig:imbh}
  \end{figure}

As already mentioned in Section \ref{sec:res}, there are two different regimes of BH mass buildup (see Fig. \ref{fig:imbh} and the "white space" - a region between about 1.0 to 3.0 Gyr with lowest density of lines, which marks approximately the separation between these two regimes). One starts later on in the cluster evolution (usually close to the core collapse time) with a small rate (SLOW scenario), while the other starts almost immediately from the beginning of our simulations (usually just after the cessation of NS and BH formation via supernovae explosions and formation of a dense self-gravitating BH sub-system) with a very high rate of mass buildup (FAST scenario). As discussed in detail later on in this section, the FAST scenario requires that a dense and massive BH subsystem forms shortly after the most massive stars explode as SN.  The SLOW scenario, on the other hand, requires that all BHs formed during SN explosions are ejected from the cluster, such that the seed BH forms later due to dynamical interactions. A substantial buildup of BH mass is found in runs both with and without mass fallback during SN explosions \citep{Belczynskietal2002}. Only a small fraction (about 20\%) of all models considered here show a substantial buildup of BH mass and subsequent IMBH formation. In general, the process of IMBH formation is highly stochastic.  Upon inspecting our simulations, we find the following trends for IMBH formation: the larger the initial cluster mass, the higher the probability of IMBH formation; the larger the initial cluster concentration, the higher the probability of IMBH formation; the larger the initial cluster concentration, the earlier and faster an IMBH will form.  

In what remains of this paper, we address the following questions:

\begin{enumerate}
\item What are the mechanisms responsible for IMBH formation? - Section \ref{sec:imbh}
\item How is it possible to form an IMBH so late in the cluster evolution? - Section \ref{sec:imbh}
\item What are the observational (direct and indirect) signatures of the presence of an IMBH in GCs? - Section \ref{sec:obs}
\end{enumerate}

The simulations presented in Fig. \ref{fig:imbh} (solid lines) were carried out with the assumption that 100\% of the mass of a star colliding with a BH is accreted onto the BH, and after the collision the size of the final object is three times larger than the sum of the radii of colliding stars (Standard accretion case). The sticky-star approximation is the standard assumption applied in the McScatter interface to the BSE code \citep{Heggieetal2006} and in the Fewbody code \citep{FCZR2004}. Additionally, the size of a star post-collision is set to three times the radius of a normal MS star with the mass of the collision product, which is the default assumption in the Fewbody code. The process of mass accretion of any type of star onto the BH is very complicated (e.g. jets, strong radiation heating the gas, etc. - all of them not included in simulations presented in this paper) and recent simulations suggest that less than 50\% of the incoming star mass can be directly accreted onto the BH \citep[see e.g.][]{GuillochonRamirez-Ruiz2013,Kyutokuetal2015,Shiokawaetal2015}. \citet{MetzgerStone2015} argue that the accreted mass fraction can be as small as 10\%, and even as small as 1\%. Such small fractions for the accreted mass are debatable given very short time scales required for SMBH formation in the early Universe. Notwithstanding, we stress that if the BH accretion efficiency is much lower than assumed in this paper, then no IMBH should form within a Hubble Time.  The exact efficiency limit is difficult to quantify given the \textit{stochasticity} of IMBH formation in our simulations.  We caution that the efficiency of IMBH formation is sensitive to the maximum initial BH mass, which is determined both by the maximum IMF mass and the amount of mass lost due to stellar evolution. These issues will be the subject of future work. In the next set of simulations, to test the IMBH formation scenario in much less favourable conditions, we therefore weaken the standard assumptions - only 25\% (and even 10\% for a few cases) of the mass of the incoming star was accreted and the size of the final merged object was just the sum of the radii of the colliding stars (Reduced accretion case). As is clear from Fig. \ref{fig:imbh} (see the dashed lines), IMBHs still form. As expected for the Reduced accretion case, the IMBH formation process is less efficient, but still two main scenarios, SLOW and FAST, are present. Typically, however, the final IMBH masses are lower in the Reduced accretion case.  

\begin{figure}
{\includegraphics[height=11cm,angle=270,width=8.5cm]{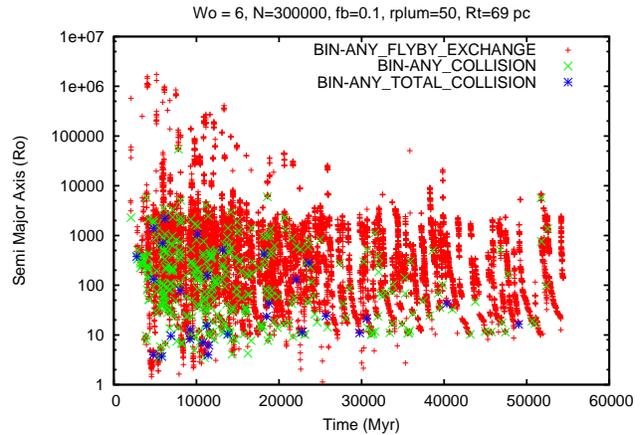}}
    \caption{Time evolution of the semi-major axis for binaries containing an IMBH in the SLOW scenario. The model parameters are given in the figure title. Symbols: plus - flybys or exchanges during binary interactions, cross - collisions during binary interactions in which the binary is preserved, star -\textit{total} collisions, where all dynamically interacting stars collide to form a single object.}
\label{fig:slow_a_bin}
  \end{figure}

\begin{figure}
{\includegraphics[height=11cm,angle=270,width=8.5cm]{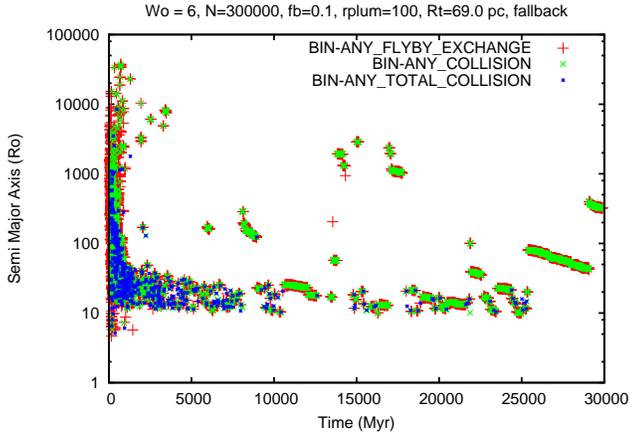}}
    \caption{Time evolution of the semi-major axis for binaries containing an IMBH in the FAST scenario. The model parameters are given in the figure title. Symbols: plus - flybys or exchanges during binary interactions, cross - collisions during binary interactions in which the binary is preserved, star - \textit{total} collisions, where all dynamically interacting stars collide to form a single object.}
\label{fig:fast_a_bin}
  \end{figure}

\begin{figure}
{\includegraphics[height=11cm,angle=270,width=8.5cm]{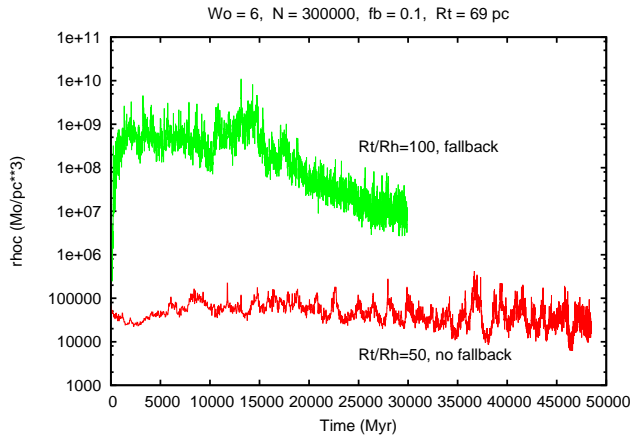}}
    \caption{Time evolution of the central cluster density in the SLOW and FAST scenarios for IMBH formation. The model parameters are given both in the figure title, and within the figure itself.}
\label{fig:rho}
  \end{figure}

Importantly, accreting BHs almost always have a binary companion (whatever its nature). Figs. \ref{fig:slow_a_bin} and \ref{fig:fast_a_bin} show the time evolution of the semi-major axes of binaries containing an IMBH in the SLOW and FAST scenarios, respectively. In the SLOW scenario, there are clearly visible binary evolution patterns (see the lines of plus signs extending nearly perpendicular to the time axis) - shrinkage of the binary semi-major axis (binary hardening). This is due to dynamical interactions with other binaries and single stars, as well as gravitational wave radiation (GR). Binary mergers due to GR are rare. More frequent are collisions connected with binary dynamical interactions (see Fig. \ref{fig:imbh_mass_slow}). In the FAST scenario, binary dynamical interactions are primarily connected with collisions or \textit{total} collisions due to the extremely high densities (see Fig. \ref{fig:rho}), small binary semi-major axis (see Fig. \ref{fig:fast_a_bin}) and very large IMBH masses (see Fig. \ref{fig:imbh}). The evolution of the binary semi-major axis displayed by the short lines of cross signs extending parallel to the time axis, visible in Fig. \ref{fig:fast_a_bin}, is connected with collisions, not with flybys. There are two kinds of dynamical collisions in binary-single and binary-binary interactions: the collision of an incoming star with one binary component (which preserves the binary), and the collision of all interacting stars three or four into one object (i.e. \textit{total} collision). During a \textit{total} collision, there is a sequence of subsequent collisions of a puffy star formed in the first collision with other interacting stars. For such an event to occur, at least one star must not be an evolved compact star (WD, NS or BH), otherwise the collision probability is extremely small. Once again, we stress that within the framework of the Fewbody code, any dissipational processes are not taken into account. So, there are no GR-mediated mergers during dynamical interactions. These can only happen via binary evolution, which is taken care by the BSE code. The \textit{total} collision events are crucial for the IMBH formation. Because of these events, an IMBH binary (when its mass is still low) can avoid hardening to the point of being able to escape from the cluster, which can occur if it receives a strong recoil during a subsequent dynamical interaction. So, the IMBH remains in the system and steadily grows in mass.  

\begin{figure}
{\includegraphics[height=11cm,angle=270,width=8.5cm]{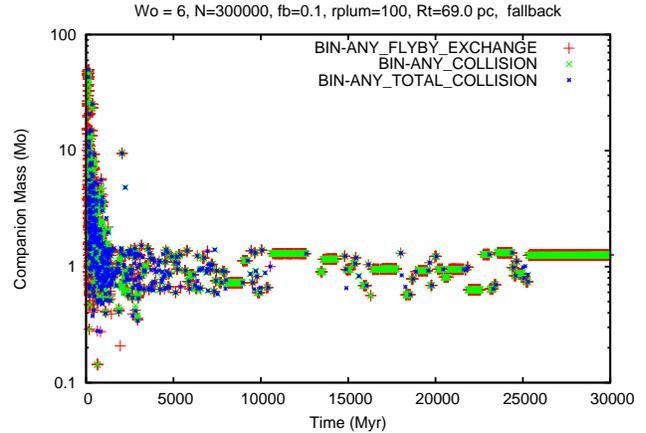}}
    \caption{Companion masses for binaries containing an IMBH and participating in dynamical interactions in the FAST scenario. The model parameters are given in the figure title. Symbols: plus - flybys or exchanges during binary interactions, cross - collisions during binary interactions in which the binary is preserved, star - \textit{total} collisions, where all dynamically interacting stars collide to form a single object.}
\label{fig:fast_m_bin}
  \end{figure}

In Fig. \ref{fig:fast_m_bin}, we show the masses of IMBH binary companions. As is clear, in the FAST scenario, the early fast increase of the IMBH mass is connected with the formation of a massive BH subsystem in the central cluster regions, and the subsequent dynamical collisions and \textit{total} collisions of massive stellar mass BHs and massive binaries containing BHs\footnote{The strong recoils due to gravitational radiation upon the mergers of BHs is not considered and quantified here. The inclusion of this effect can slow down IMBH mass buildup, but not prevent IMBH formation.}.  In these models, a large number of BHs remain in the system due to a reduction in the imparted kick velocity that follows from the assumption of significant mass fallback onto the BHs. These BHs have masses up to $60 M_{\odot}$, and are usually the companions of binaries with accreting IMBHs.  After about 2 Gyr, these very massive objects are no longer present in the system. They end up being removed by, or colliding with, the IMBH due to dynamical interactions. At this time, the IMBH companion masses in the FAST scenario are similar to those in the SLOW scenario. These companions have masses around $1 M_{\odot}$, which are characteristic of the most massive objects remaining in the system at this time, such as NSs, WDs, MS stars and even the collision products of unevolved stars. So, after all stellar mass BHs have been removed from the system, the rate of IMBH mass growth substantially decreases, becoming more or less similar to what is seen in the SLOW scenario. 

\begin{figure}
{\includegraphics[height=11cm,angle=270,width=8.5cm]{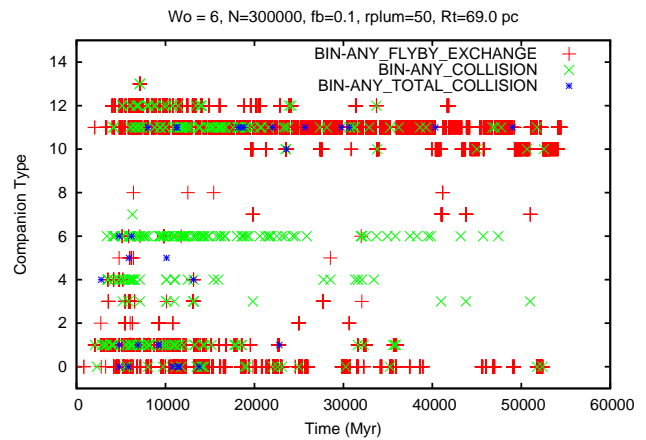}}
    \caption{Companion types for binaries containing an IMBH that participate in dynamical interactions in the SLOW scenario. The model parameters are given in the figure title. Symbols: plus - flybys or exchanges during binary interactions, cross - collisions during binary interactions in which the binary is preserved, star - \textit{total} collisions, where all dynamically interacting stars collide to form a single object.}
\label{fig:slow_type_bin}
  \end{figure}

\begin{figure}
{\includegraphics[height=11cm,angle=270,width=8.5cm]{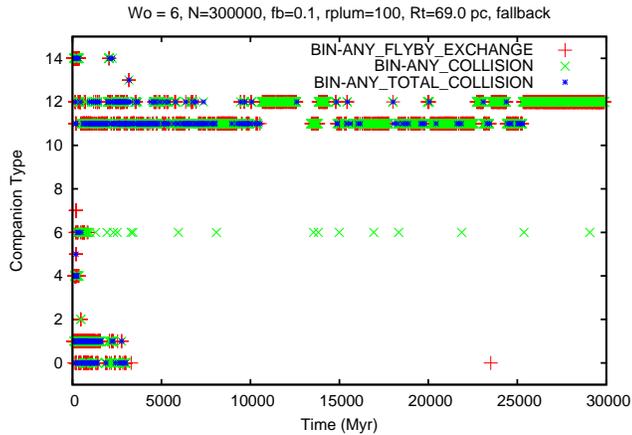}}
    \caption{Companion types for binaries containing an IMBH that participate in dynamical interactions in the FAST scenario. The model parameters are given in the figure title. Symbols: plus - flybys or exchanges during binary interactions, cross - collisions during binary interactions in which the binary is preserved, star - \textit{total} collisions, where all dynamically interacting stars collide to form a single object.}
\label{fig:fast_type_bin}
  \end{figure}

The conclusions reached in the previous paragraph are further supported by the data presented in Figs. \ref{fig:slow_type_bin} and \ref{fig:fast_type_bin}, where the types of IMBH binary companion stars are shown. In the SLOW scenario, events involving any type of unevolved companion stars are spread out over the whole cluster evolution.  In the FAST scenario, however, they are present only at the beginning of our simulations, with a few additionally events involving asymptotic giant branch stars occurring later on in the cluster evolution. In both cases, the most frequent type of dynamical interaction involves an IMBH binary with a WD companion. Interactions with IMBH binaries containing NSs and BHs are in general much more rare, but occur more frequently in the FAST scenario (near the beginning of the cluster evolution). In the FAST scenario, \textit{total} collisions occur with the greatest frequency, due to the large cluster density (see Fig. \ref{fig:rho}) and the small IMBH binary semi-major axis (see Fig. \ref{fig:fast_a_bin}). The IMBH mass becomes so high that gravitational focusing efficiently channels incoming stars/binaries to drift very close to the IMBH, such that strong interactions occur regularly.  This serves to trigger additional merger events with the IMBH. 

As it is clear from Fig. \ref{fig:rho}, the central cluster densities reached in the SLOW scenario are not very high - only $10^5 M_{\odot}/pc^3$ (similar to the observed densities of many Galactic GCs). Thus, IMBH formation does not require any special sets of conditions. The situation is different in the FAST scenario. The densities required for significant IMBH mass buildup to occur are very high, greater than $10^8 M_{\odot}/pc^3$. These extremely high densities are needed when BHs form a bound and very dense subsystem in the cluster centre - collisions of binaries containing BHs must be more efficient than the removal of the most massive BHs from the cluster due to strong recoils during dynamical interactions. The central escape velocity is on the order of a few hundred km/s. Such high densities and central escape velocities are unlikely to be reached in the GCs observed in the Milky Way \footnote{The most dense Galactic GCs have central density around $10^6 M_{\odot}/pc^3$ - see \citet[][updated 2010]{Harris1996}}.  These densities can be much higher in nuclear star clusters (NSCs). Could the FAST scenario for IMBH formation discussed in this paper occur commonly in the NSCs of low-mass galaxies?  

Based on the results presented in this section, the new scenario for IMBH formation proposed in this paper can be summarised as follows:

\begin{itemize}
\item There are two possible variants of the IMBH formation scenario: SLOW and FAST.
\begin{itemize}
\item SLOW scenario - either a single BH is left after the early phase of SN explosions, or a single BH is formed via mergers or collisions during dynamical interactions;
\item FAST scenario - several dozen BHs remain in the system after the early phase of SN explosions, and form a dense central subsystem.  The central density must be extremely high (greater than  $10^8 M_{\odot}/pc^3$) for an IMBH to form. Alternatively, all BHs are quickly and efficiently removed from the system via dynamical interactions. If at least one remains, then the SLOW scenario is followed\footnote{Interestingly, a dense, gravitationally bound BH subsystem can form without assuming any reduction (due to mass fallback) in the standard prescription for SN kicks \citep{Belczynskietal2002}. A substantial number of BHs can be retained in the system post-kick due to the large escape velocity, which can be on the order of about 200 km/s for such dense clusters.}.
\end{itemize}
\item Next, the formation of a BH-binary (i.e. a BH in a binary with any type of star as its companion) forms via a three-body interaction. The BH is the most massive object in the cluster, so there is a high probability that it will form or it will be exchanged into a binary. 
\item Dynamical interactions with other binaries and stars:
\begin{itemize}
\item orbit tightening leading to mass transfer from MS/RG/AGB companions;
\item exchanges and collisions, leaving the binary in tact;
\item \textit{total} collisions during dynamical interactions or mergers induced by the emission of gravitational waves - in this case, the binary is destroyed and only a BH is left. The single BH then forms a new binary via another three-body interaction, which is free to undergo subsequent dynamical interactions with other single and binary stars, and the process repeats.  In this way, the BH mass steadily increases.
\end{itemize}
\end{itemize}
It is worth noting that the presented scenarios for IMBH formation in GCs, in particular the SLOW scenario, do not require any specific conditions, unlike other scenarios proposed in the literature. IMBH formation occurs solely via binary dynamical interactions and mass transfer in binaries.

Having discussed the two regimes of formation scenarios for IMBHs and their conditional requirements, we now turn our attention to possible observational signatures of the presence of IMBHs in star clusters.

\section{Observational Signatures of IMBHs in Globular Clusters}\label{sec:obs} 

As already stated in Section \ref{sec:int}, IMBHs are only observable via indirect means. What is observed is their influence on the host cluster structure or radio and X-ray emissions connected with accretion processes. In the future, it may also be possible to observe GR from mergers of compact objects with an IMBH. In this Section, we discuss possible observational signatures of an IMBH.  

\subsection{Surface Brightness and Velocity Dispersion Profiles}\label{sec:sbp} 

Theoretical investigations and numerical simulations suggest that, if an IMBH is present in the centre of a GC, the velocity dispersion profile (VDP) should rise toward the cluster centre, and the surface brightness profile (SBP) should be shallow with a large core radius. This suggests that the best candidates to harbour IMBHs are massive GCs with large core radii, shallow central SBPs and centrally rising VDPs. However, as shown by \citet{Bianchinietal2015}, when using integrated light, the observed VDP can be heavily biased by the presence of a few very bright stars in the cluster core. This would obscure any underlying observational signature of an IMBH.

\begin{figure}
{\includegraphics[height=11cm,angle=270,width=8.5cm]{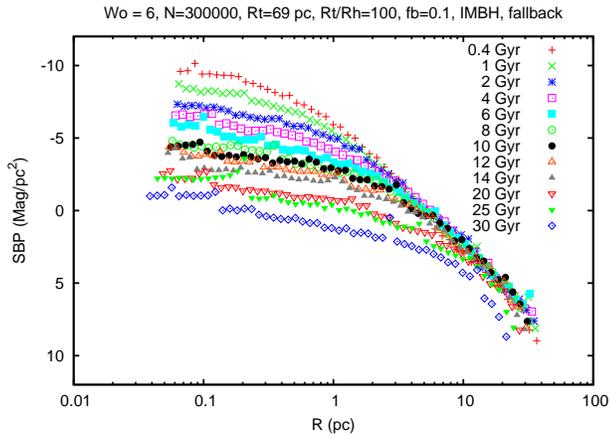}}
    \caption{Surface brightness profiles (for the times given in each inset) in the FAST scenario. An IMBH begins forming at the very beginning of the cluster evolution. The model parameters are given in the figure title.}
\label{fig:sbp_imbh}
  \end{figure}

\begin{figure}
{\includegraphics[height=11cm,angle=270,width=8.5cm]{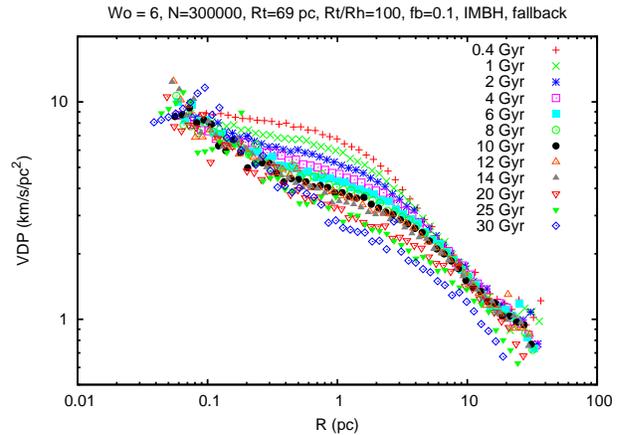}}
    \caption{Velocity dispersion profiles (for the times given in each inset) in the FAST scenario. VDP is computed only for stars which $M_V < 6$ mag. An IMBH begins forming at the very beginning of the cluster evolution. The model parameters are given in the figure title.}
\label{fig:vdp_imbh}
  \end{figure}

\begin{figure}
{\includegraphics[height=11cm,angle=270,width=8.5cm]{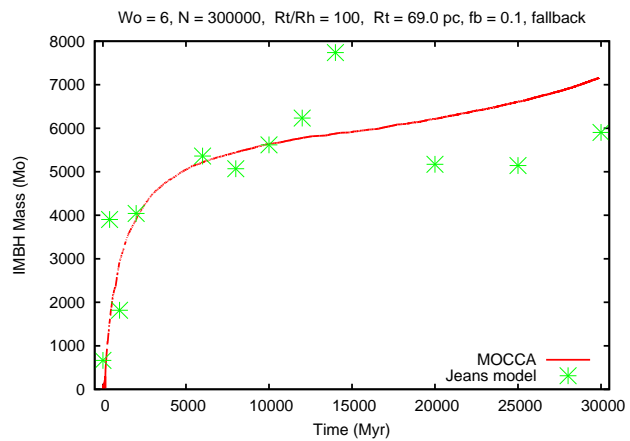}}
    \caption{The time evolution of the IMBH mass for a particular MOCCA simulation, found from fitting Jeans' model \citep{Lutzgendorfetal2013} to the simulated SBP and VDP. The model parameters are given in the figure title.}
\label{fig:fit_jeans}
\end{figure}

In Figs. \ref{fig:sbp_imbh} and \ref{fig:vdp_imbh} are shown the SBP and VDP for a particular model featuring IMBH formation. A flat SBP with a large core radius and a steeply rising VDP toward the cluster centre are clearly visible. The IMBH begins accreting mass almost immediately, eventually increasing its mass to 7000 $M_{\odot}$. The simulated VDPs and SBPs were used to fit the IMBH mass using Jeans' model \citep[see][for a detailed description of the method]{Lutzgendorfetal2013}. Fig. \ref{fig:fit_jeans} shows that the resulting fit to the MOCCA data is quite good. The best agreement is found between about 4 to 12 Gyr after t $=$ 0, at which time the IMBH mass is already substantial, albeit a small fraction of the total cluster mass. At earlier times, the IMBH is quickly growing in mass.  At even later times (i.e. $>$ 12 Gyr), the IMBH mass becomes a substantial fraction of the total cluster mass (the IMBH mass is about 15\% of the total cluster mass at t $=$ 20 Gyr). Hence, at early and late times, the applied fitting method reaches its limits: at early times the IMBH mass is too small to yield a strong imprint on the observed SBPs and VDPs, whereas at late times the IMBH mass dominates the central cluster potential. The fact that the MOCCA code is able to correctly reproduce the cluster structure in the vicinity of an IMBH is somewhat surprising.  This is because MOCCA was not designed to model the cluster evolution on time scales relevant to a massive IMBH. Hence, the successful application of the MOCCA code to IMBH formation highlights a number of strengths inherent to the Monte Carlo method for cluster evolution that we now understand.  Nevertheless, models that feature IMBHs must be inspected with caution, since their presence can be masked by other dynamical effects.

\begin{figure}
{\includegraphics[height=11cm,angle=270,width=8.5cm]{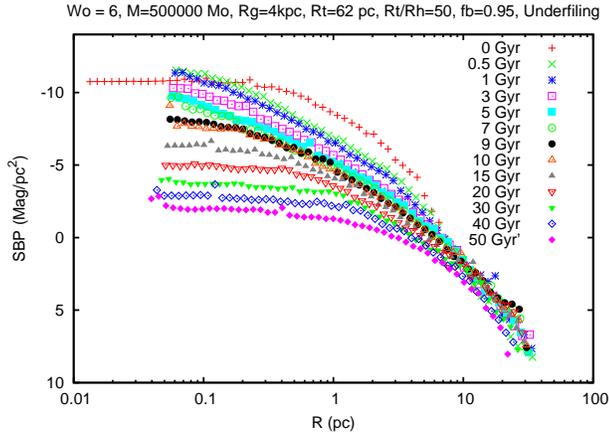}}
    \caption{Surface brightness profiles (for the times given in each inset) in the SLOW scenario. An IMBH begins forming at a cluster age of about 10 Gyr. The model parameters are given in the figure title.}
\label{fig:sbp_noimbh}
  \end{figure}

\begin{figure}
{\includegraphics[height=11cm,angle=270,width=8.5cm]{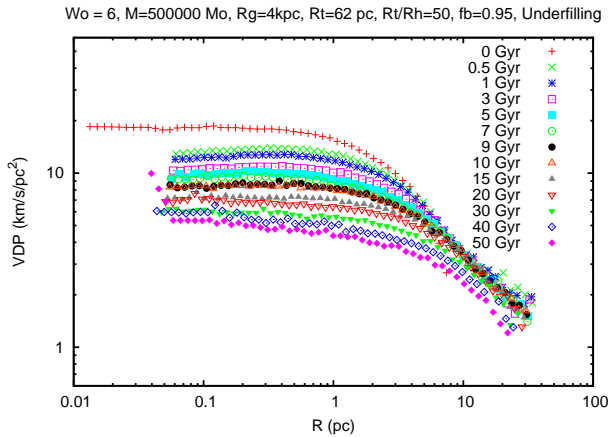}}
    \caption{Velocity dispersion profiles (for the times given in each inset) in the SLOW scenario. VDP is computed only for stars which $M_V < 6$ mag. An IMBH begins forming at a cluster age of about 10 Gyr. The model parameters are given in the figure title.}
\label{fig:vdp_noimbh}
  \end{figure}

For example, some models in which an IMBH forms lack a centrally rising VDP and/or a shallow central SBP. One such example is shown in Figs. \ref{fig:sbp_noimbh} and \ref{fig:vdp_noimbh}. In this model, IMBH mass growth begins about 10 Gyr into the cluster evolution.  
Up until this time, the SBP shows clear signs of core collapse, namely a centrally rising profile. Later on, the SBP profile begins to flatten. Meanwhile, the VDP does not show the characteristic structure of either a collapsing core or a core hosting an IMBH. Instead, the VDP first begins to decrease toward the cluster centre, and then becomes flat. Only, at a cluster age of about 
50 Gyr is there a clear indication of a steeply increasing VDP toward the cluster centre.

\begin{figure}
{\includegraphics[height=11cm,angle=270,width=8.5cm]{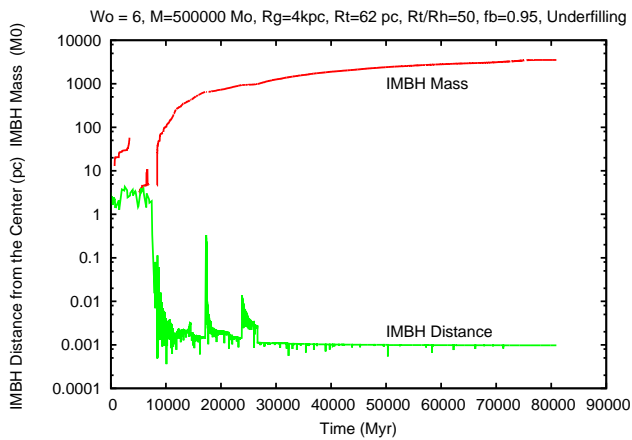}}
    \caption{The distance of the IMBH from the cluster centre along with the IMBH mass as a function of time in the SLOW scenario. The model parameters are given in the figure title.}
\label{fig:imbh_r_m}
  \end{figure}

To better understand these effects, we turn our attention to the motion of the IMBH relative to the cluster centre. MOCCA treats IMBHs as ordinary objects, albeit extremely massive ones. They are not pinned to the cluster centre, as is assumed in Jeans' model. In Fig. \ref{fig:imbh_r_m} we show the position of an IMBH with respect to the cluster centre as a function of time. Clearly, the motion of the IMBH is significant, particularly for IMBH masses less than a dozen solar masses. There are visible spikes in this motion connected with strong dynamical interactions between the IMBH and surrounding objects. We now know (e.g. Spurzem, Diploma thesis University of Goettingen 1983, personal communication, \citet[][and references therein]{ZhongBS2014}) that motion of IMBHs and SMBHs relative to the centre of mass of their host can destroy a Bahcall-Wolf cusp \citep{BahcallWolf1976}, which is ultimately responsible for the steeply rising central VDP. We stress that within the framework of the MOCCA code, the off-centre positions of massive IMBHs can alter the structure of the underlying cluster potential, and hence influence the orbits of cluster stars. Based on the model presented in Figs. \ref{fig:sbp_noimbh}, \ref{fig:vdp_noimbh}, \ref{fig:imbh_r_m} and an analysis of other models, the presence of an IMBH will show a clear signal in the VDP for IMBH masses larger than about $1000 - 2000 M_{\odot}$. 

\subsection{Electromagnetic and Gravitational Radiation from Accreting IMBHs}\label{sec:gr}

\begin{figure}
{\includegraphics[height=11cm,angle=270,width=8.5cm]{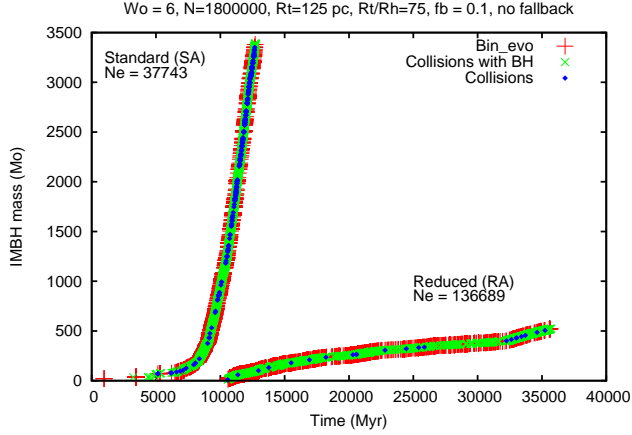}}
    \caption{The time evolution of the IMBH mass for models with Standard (SA) and Reduced (RA) mass accretion rates in the SLOW scenario. The model parameters are given in the figure title. Accretion events due to binary stellar evolution, collisions and hyperbolic collisions are depicted by different symbols: plus signs, crosses and dots, respectively. Also listed is the total number of IMBH interactions (Ne). The numbers of particular types of interactions, and the total mass gained by the IMBH in each such dynamical interaction or binary evolution-mediated mass transfer event, are as follows: Mergers with IMBH (RA) - $N = 6049$, $M = 277.1 M_{\odot}$, Collisions with IMBH in binary (RA) - $N = 1175$, $M = 236.1 M_{\odot}$, Hyperbolic Collisions (RA) - $N = 23$, $M = 8.1 M_{\odot}$, Mergers with IMBH (SA) - $N = 3490$, $M = 354.1 M_{\odot}$, Collisions with IMBH in binary (SA) - $N = 3110$, $M = 2915.7 M_{\odot}$, Hyperbolic Collisions (SA) - $N = 341$, $M = 254.0 M_{\odot}$. The numbers of different GR-related events are as follows: SA case - IMBH-WD mergers  = 87, IMBH-NS mergers = 0, IMBH-BH mergers = 0, RA case - IMBH-WD mergers  = 159, IMBH-NS mergers = 12, IMBH-BH mergers = 2.}
\label{fig:imbh_mass_slow}
  \end{figure}

In Fig. \ref{fig:imbh_mass_slow}, we show the time evolution of the IMBH mass in the SLOW scenario for models with Standard and Reduced accretion rates (see the definitions of Standard and Reduce accretion in Section \ref{sec:imbh}). In the figure caption are given the numbers and properties of mass-transfer events onto the IMBH, either from a companion during binary evolution, or because of collisions with incoming stars. As expected, the Reduced accretion rate case shows a smaller rate of IMBH mass buildup relative to the Standard case. For the Standard case, the amount of mass accreted by the IMBH via collisions during dynamical interactions is much larger than for binary evolution-mediated mass-transfers or mergers, despite the fact that the number of times each event occurs is roughly the same. For the Reduced case, the amount of mass accreted by the IMBH is roughly the same for both mechanisms, despite the fact that the number of binary evolution-mediated mass-transfers or mergers strongly outweighs the number of collisions during dynamical interactions. The rate of IMBH mass growth due to hyperbolic collisions is much larger in the Standard case than in the Reduced case.  This is because the formation of binaries in three-body interactions becomes less probable with increasing IMBH mass due to the large accelerations imparted very close to the IMBH. Interestingly, mergers due to GR and binary evolution are more frequent in the Reduced case. This is connected with the fact that collisions during dynamical interactions produce far fewer \textit{total} collisions, due to the smaller size and hence cross-section of the collided object. Instead, a large number of GR-mediated mergers are observed, mainly associated with IMBH-WD mergers. Because of the large masses of IMBHs, GR-induced mergers can be effective for semi-major axes as large as several hundred solar radii. In general, collisions, mergers with or mass transfer onto an IMBH are potentially observable via the associated electromagnetic and/or gravitational radiation. However, the MOCCA and BSE codes are unable to accurately quantify the duration of these events. Obviously, this presents a significant challenge in estimating the probability of observing them.  

In the FAST scenario (not shown), hyperbolic collisions are the dominant mechanism for IMBH mass growth provided the IMBH is sufficiently massive (larger than 3000 - 5000 $M_{\odot}$).  For smaller IMBH masses, the processes described in the previous paragraph apply instead. The large number of hyperbolic collisions is connected with the extreme central densities and IMBH masses. In the vicinity of a massive IMBH, binary formation during a three-body interaction is unlikely, and gravitational focusing increases the probability of direct collisions. There are also hundreds of binary mergers due to GR.  

\subsection{Hypervelocity Escapers}\label{sec:esc} 

In recent years, a large number of hypervelocity star (HVS) candidates have been reported \citep[just to list the most resent papers:][and references therein]{BrownGK2014,Lietal2012,Palladinoetal2014,Zhengetal2014}. More than 50 stars have now been classified as HVSs with velocities larger than 400 km/s. HVSs can obtain their large velocities via various channels, including a variety of dynamical mechanisms and SN kicks \citep[e.g.][]{BaumgardtGP2006,Tauris2015}. The dynamical channels that create HVSs are connected with interactions between: a SMBH/IMBH and a binary, a SMBH/IMBH-SMBH binary and a (either single or binary) star, and a SMBH/IMBH and two unbound stars. The formation of IMBHs in our models create HVSs. 

\begin{figure}
{\includegraphics[height=11.0cm,angle=270,width=8.5cm]{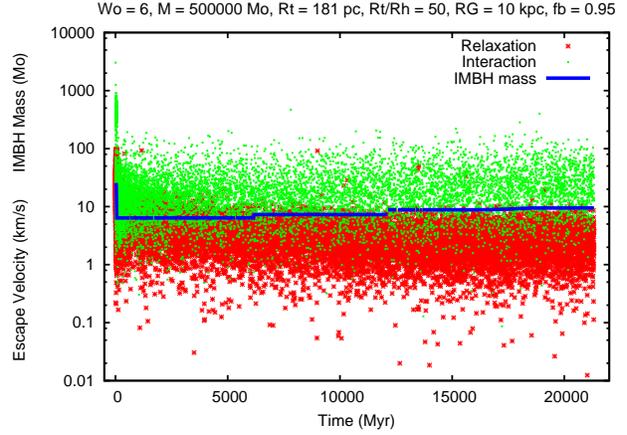}}
    \caption{The escape velocities of objects removed from the system because of relaxation (star symbol) and dynamical interactions (dot symbol) for models that do not feature IMBH formation. The line shows the mass of the most massive BH in the system.}
\label{fig:escape_no}
  \end{figure}

\begin{figure}
{\includegraphics[height=11cm,angle=270,width=8.5cm]{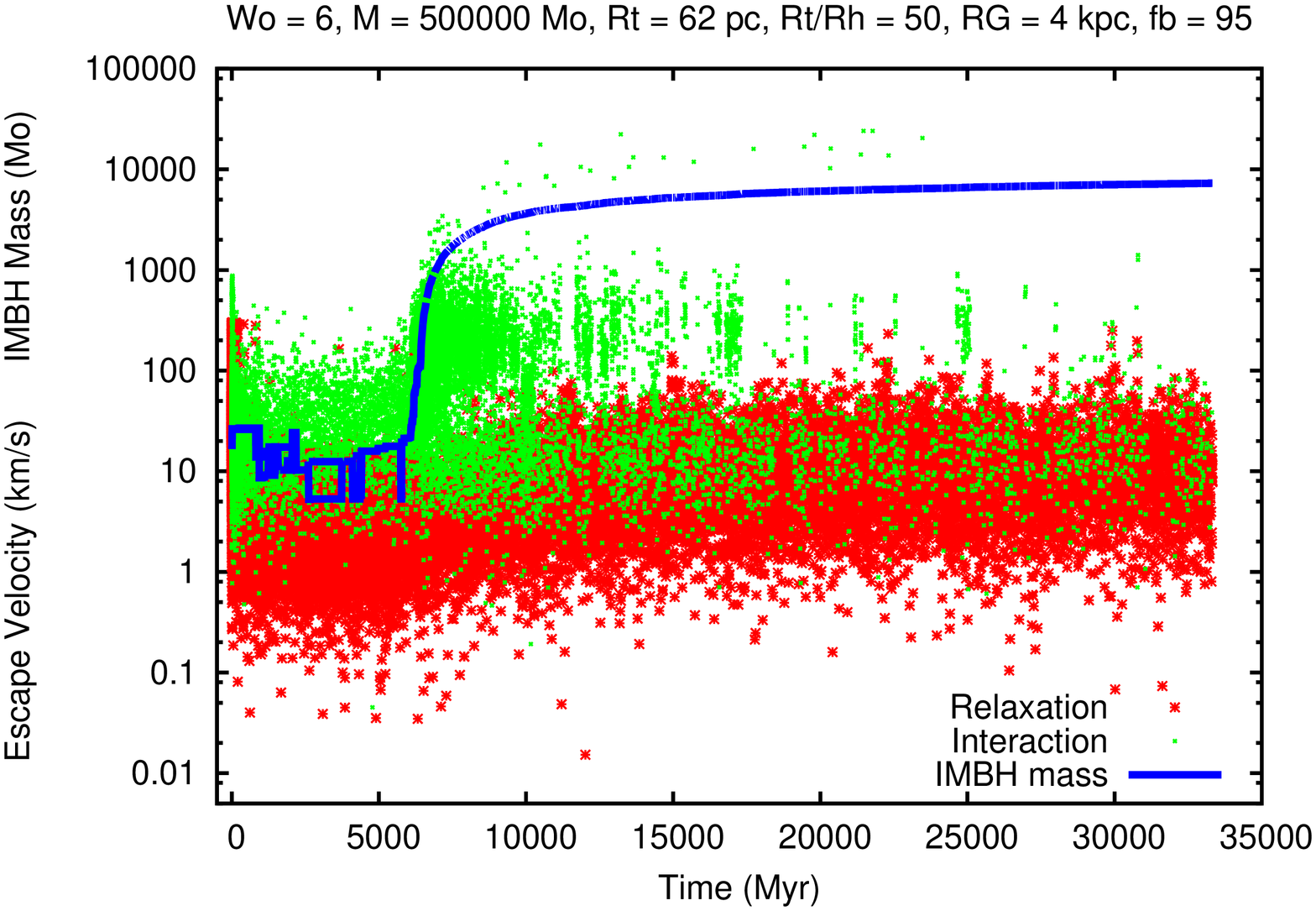}}
    \caption{The escape velocities of objects removed from the system because of relaxation (star symbol) and dynamical interactions (dot symbol) for models featuring IMBH formation - SLOW scenario. The line shows the mass of the most massive BH in the system.}
\label{fig:escape_slow}
  \end{figure}

\begin{figure}
{\includegraphics[height=11cm,angle=270,width=8.5cm]{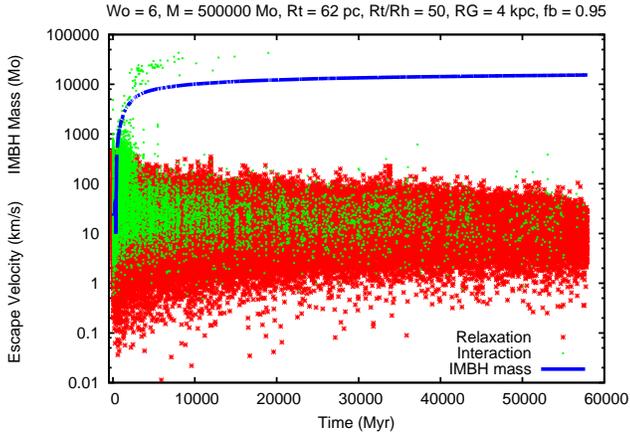}}
    \caption{The escape velocities of objects removed from the system because of relaxation (dot symbol) and dynamical interactions (dot symbol) for models featuring IMBH formation - FAST scenario. The line shows the mass of the most massive BH in the system.}
\label{fig:escape_fast}
  \end{figure}

In Figs. \ref{fig:escape_no}, \ref{fig:escape_slow} and \ref{fig:escape_fast} we show the escape velocities of objects removed from the system because of relaxation and dynamical interactions, both for models that do not feature IMBH formation and for models that do, both in the SLOW and FAST scenarios, respectively. We expect objects that escape because of relaxation to have relatively low velocities (kilometres per seconds), and objects escaping due to dynamical interactions to have relatively large velocities (dozenths or hundreds kilometres per second). Indeed, this is the case for models that do not feature IMBH formation (see Fig. \ref{fig:escape_no}). However, a more complicated picture emerges for models that feature IMBH formation (see Fig. \ref{fig:escape_slow} and \ref{fig:escape_fast}). We observe the following features: when a very massive IMBH forms with a mass greater than a few hundred - $1000 M_{\odot}$, it is able to produce relaxation escapers with velocities on the order of about 100 km/s. These large escape velocities are connected with the very large velocities of stars in the immediate vicinity of the IMBH. Clearly, the formation of IMBHs with masses larger than $100 - 200 M_{\odot}$ is connected with higher velocities of dynamical escapers. For IMBH masses up to $1000 - 2000 M_{\odot}$, a substantial number of escapers are connected with dynamical interactions involving binaries, and produce escapers with velocities up to 1000 - 2000 km/s. For larger IMBH masses, the number of dynamical escapers decreases, but their velocities increase substantially, up to 10000 km/s or more. This decrease in the number of dynamical escapers is connected with a decrease in the probability of binary formation. New binaries containing very massive IMBHs rarely form (due to \textit{total} collisions of binaries).  Consequently, strong binary dynamical interactions are rare, and hence rarely produce escapers.  

Very early on in the cluster evolution, there are also HVS escapers (NSs and BHs) connected with SN kicks.

Concluding, the presence of IMBHs in GCs is potentially indirectly observable via the presence of anomalously high numbers of HVSs with measured 3D space velocities extending back toward their former host GCs, which might be identifiable using the GAIA satellite.  Some of these should be low-mass MS stars (with masses close to the turn off mass). 

\section{Galactic Globular Clusters}\label{sec:dis}

As originally presented in \citet{Leighetal2013b} and \citet{Leighetal2015a}, the GC models discussed in this paper are able to reproduce the observed correlations between a number of global GC parameters, including the total cluster mass, concentration parameter, mass function slope and binary fraction. Here, we re-visit this exercise in order to see if models that feature IMBH formation are able to produce present-day cluster properties consistent with those of real Galactic GCs (Harris 1996, updated 2010). 

We assume that the probability of IMBH formation depends mainly on the average binary interaction probability and the escape velocity. The larger the escape velocity, the larger the number of NSs and BHs retained in the system, which opens the possibility for IMBH formation. The larger the probability for binary interactions, the larger the probability of \textit{total} collisions of binaries, which prevent massive BHs from being removed from the system via strong dynamical interactions. Hence, we ask:  Do models that feature IMBH formation occupy regions in Escape Velocity - Probability of Interaction space different from models that do not feature IMBH formation? Naively, we expect models that form an IMBH to have a high probability for binary interactions but a low probability of escape.  Conversely, models that do not form an IMBH should have a high escape probability but a low probability for binary interactions.  

\begin{figure}
{\includegraphics[height=11cm,angle=270,width=8.5cm]{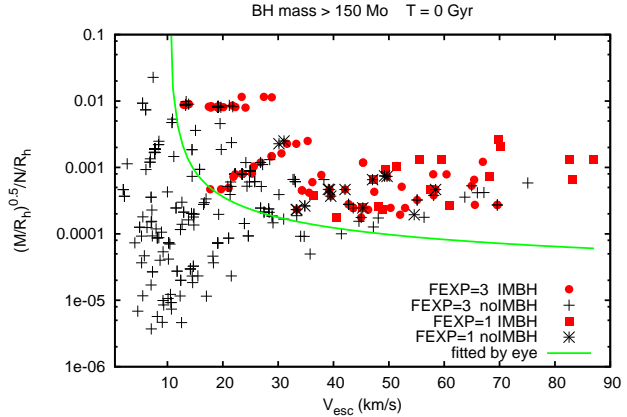}}
    \caption{The interaction probability vs escape velocity at time T=0 for all models. Only models for which BHs form with masses greater than 150 $M_{\odot}$ are considered as models containing IMBHs. The filled symbols correspond to models featuring IMBH formation, while the open symbols correspond to models without IMBH formation. The circles correspond to models that assume the standard accretion rate, while the squares correspond to models that assume the reduced accretion rate (see text for more details). The line (drawn by eye) is a statistical boundary separating models with and without IMBH formation.} 
\label{fig:imbh_t0}
  \end{figure}

\begin{figure}
{\includegraphics[height=11cm,angle=270,width=8.5cm]{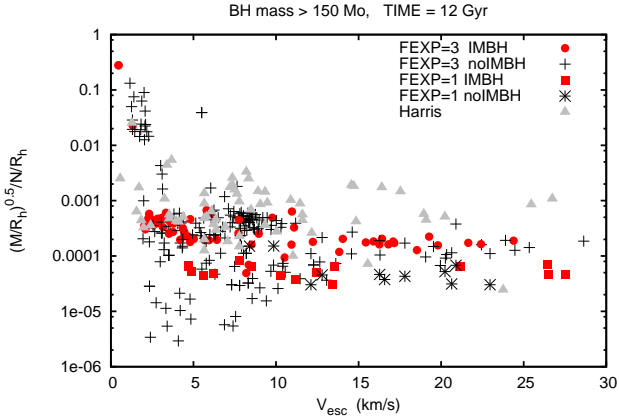}}
    \caption{The interaction probability vs escape velocity at time T=12 for all models. Only IMBHs with masses greater than 150 $M_{\odot}$ are considered. The filled symbols correspond to models featuring IMBH formation, while the open symbols correspond to models without IMBH formation. The circles correspond to models that assume the standard accretion rate, while the squares correspond to models that assume the reduced accretion rate (see text for more details). The triangles correspond to the observed GC parameters taken from the Harris catalogue \citep[][updated 2010]{Harris1996}}
\label{fig:imbh_t12}
  \end{figure}

The probability of interaction can be approximated by considering an interaction between an average binary at the soft/hard boundary and an average single star in the average cluster environment inside $R_h$. So, for binaries with binding energies equal to the average single star kinetic energy, the binary semi-major axis is $a_{s-h} \propto R_h/N$, where N is the number of stars in the system and $R_h$ is the half-mass radius. The average stellar mass is $<m>\quad= M/N$, where M is the total cluster mass. The average cluster density and velocity dispersion inside $R_h$ are $<n_h>\quad\propto N/R_h^3$ and $<V_h>\quad\propto sqrt({M/R_h})$, respectively, where $<n_h>$ is the average number density inside $R_h$ and $<V_h>$ is the average system velocity.
\begin{equation}
  P_{s-h} = A {M^{1/2}\over{R_h^{3/2} N}} \Delta T  
\end{equation}
where $A$ is a constant factor and $\Delta T$ is assumed to be 1 Myr.
Fig. \ref{fig:imbh_t0} shows all simulated models. Indeed, at T $=$ 0 there is a statistical boundary above which the probability for IMBH formation is high (filled symbols). We caution the reader that IMBH formation is a highly stochastic process. Different statistical realisations of models with identical initial conditions can yield very different evolutions with or without IMBH formation. The boundary line in Fig. \ref{fig:imbh_t0} is drawn by eye. Models with Standard and Reduced mass accretion only partially overlap.  For the Reduced case, an IMBH is likely to form in models with escape velocities upwards of about 30-40 km/s.  For the Standard case, an IMBH can form in models with escape velocities upwards of only 10-20 km/s  We redo Fig. \ref{fig:imbh_t0} at time T=12 Gyr in Fig. \ref{fig:imbh_t12} (using the present-day observed and modelled cluster parameters).  Here, the picture is less clear. Models with and without IMBH formation overlap. We note some Galactic GCs occupy regions in Escape Velocity - Probability of Interaction space for which MOCCA will produce an IMBH (see Fig. \ref{fig:imbh_t12}). These include Omega Cen, 47Tuc, M22 and NGC6293. 

Concluding, it is worth mentioning that models with reduced natal kicks (because of mass fallback) for BHs may contain a substantial number of stellar mass BHs even after a Hubble time. The number of retained BHs depends on the total cluster mass and concentration, and therefore the cluster half-mass relaxation time \citep{BreenHeggie2013}. The longer the half-mass relaxation time, the larger the number of retained BHs. This means that stellar mass BHs are most likely to be observed in massive and low concentration (large) GCs, instead of massive and dense GCs.  Also, simulations of dense and massive GCs show that NSs and BHs can form in substantial numbers because of hyperbolic collisions and binary collisions during binary dynamical interactions later on in the cluster evolution, not only because of supernovae explosions early on in the cluster evolution. Interestingly, IMBH formation suppresses the production of binaries with NS or BH companions, and even the formation of NSs and BHs themselves. This is because these objects (WDs, NSs and BHs) tend to be quickly removed from the system via dynamical interactions with the IMBH. These are the most massive objects in the cluster, so that the probability of their being involved in dynamical interactions is the greatest.

\section{Conclusions}\label{sec:conc} 

The MOCCA code (Giersz et al 2013) is at present one of the most advanced numerical codes for stellar dynamical simulations, capable of following the evolution of real star clusters in detail comparable to N-body simulations, but orders of magnitude faster (about a day for $N = 2 \times 10^6$). The ability of the code to model the evolution of specific star clusters (M4, NGC6397, 47Tuc and M22) and to reconstruct the observed correlations of various cluster parameters have been demonstrated in previous papers. The new scenario for IMBH formation in GCs discussed in this paper came as a byproduct of two earlier projects \citep{Leighetal2013b,Leighetal2015a} which aimed to explain the observed correlations between cluster mass and binary fraction and between cluster concentration and mass function slope. This new scenario does not require any specific conditions. IMBHs form solely via binary dynamical interactions and mass transfer in binaries, with the latter playing an especially important role by inducing collisions. There are two kinds of dynamical collisions: collisions involving one binary component (which preserve the binary), and \textit{total} collisions involving all interacting stars which collide to form a single object. \textit{Total} binary collisions are especially crucial for IMBH formation.  This is because IMBH binaries (when their masses are still low) can avoid hardening to the point of being able to escape from the cluster.  This occurs if the IMBH binary receives a strong recoil during a dynamical interaction. Instead, the IMBH remains in the system and, consequently, is able to steadily grow in mass.  

It is worth noting that to observe growth of IMBH mass one needs clusters which consists of hundreds of thousands stars and are fairly concentrated. Such conditions are difficult (and for a while still infeasible) for direct N-body simulations and therefore the process of IMBH formation was not observed yet in N-body simulations. The one exception are N-body simulations of dense open clusters done by Olczak (private communication). He observed formation of BH with masses of about 200 $M_{\odot}$ during the first a few hundred Myr of cluster evolution. Such BH mass growth was also seen in MOCCA simulations for the same initial model. Also, Monte Carle models of star clusters simulated so far were not enough initially concentrated and massive.

We find two different regimes of BH mass growth - the SLOW and FAST scenarios. The SLOW scenario is initiated at later times in the cluster evolution, has a small accretion rate and requires modest cluster densities of around $10^5 M_{\odot}/pc^3$.  The FAST scenario is initiated practically from the very beginning of our simulations, has a very high accretion rate and requires extreme densities of around $10^8 M_{\odot}/pc^3$. Generally, only a fraction (about 20\%) of all models considered here show significant BH mass growth and IMBH formation. Thus, the process of IMBH formation is highly stochastic. The larger the initial cluster concentration, the larger the probability of IMBH formation and the earlier and faster the IMBH is formed.

IMBHs can only be observed indirectly via their influence on the host cluster structure, populations of particular types of objects (including binaries, WDs, NSs, BHs or hypervelocity star escapers) or radio, X-ray and gravitational wave emissions connected with accretion processes and/or the mergers of compact objects. 

Below we provide our main conclusions, which summarise the paper.
\begin{itemize}
\item NSs and BHs can form (in substantial numbers) in the course of star cluster evolution due to dynamical interactions (collisions and binary interactions). The number of NSs/BHs (single or in binaries) retained in the system is much larger in systems without an IMBH. If an IMBH is present, all NSs and BHs are quickly removed from the cluster via dynamical interaction with the IMBH;
\item If the system density is on the order of $10^5 M_{\odot}/pc^3$ and only one stellar mass BH is present, it is possible that BH mass growth will proceed due solely to dynamical interactions and mass transfer from binary companions. In this SLOW scenario, IMBH formation happens at late times in the cluster evolution, typically during the post core-collapse phase of evolution. The SLOW scenario for IMBH formation is more probable than the FAST scenario;
\item If the system density is extremely high (greater than about $10^8 M_{\odot}/pc^3$), a gravitationally bound BH subsystem can form early on in the cluster evolution.  Subsequently, an IMBH can rapidly form via dynamical interactions between single and binary BHs. In this FAST scenario for IMBH formation, the large cluster density (and escape velocity) serves to drive the growth of the most massive BH in the system. The FAST scenario might occur more commonly in the NSCs of low-mass galaxies than in GCs;
\item The process of BH mass growth and finally IMBH formation is highly stochastic. The rate of IMBH mass growth depends sensitively on the cluster density. The larger the density, the higher the rate;
\item There are frequent but brief episodes of mass transfer from binary companions onto an IMBH, subsequent mergers and collisions with incoming stars. Therefore, it could be possible to observe such events in action via X-ray and/or GR emissions.  However, our results suggest that the probability of such an observation is very low and difficult to estimate given the uncertain nature of the initial conditions and the stochastic nature of the evolution;
\item We observe a centrally rising VDP and shallow SBP profiles with large core radii in a number of simulations with IMBH formation. For these models, we have checked that the technique for IMBH mass extraction applied in \citet{Lutzgendorfetal2013} recovers reasonably well the IMBH masses in our simulations.  However, a significant fraction of models with IMBH formation yield VDP and SBP profiles for which Jeans' model fitting cannot accurately recover the IMBH mass. This is connected with the motion of the IMBH relative to the cluster centre, which ultimately erases any underlying Bahcall-Wolf cusp and centrally rising VDP profile;   
\item Very massive (greater than a few $1000 M_{\odot}$) IMBHs with binary companions can produce hypervelocity escapers during dynamical interactions, with velocities greater than a few 1000 km/s. Most of these escapers are NSs and WDs, but some are MS stars with masses near that of the turn off. Observations of these hypervelocity stars could provide unambiguous evidence of the presence of an IMBH;
\item The present-day properties of models featuring IMBH formation are consistent with the observed properties of some Galactic GCs, e.g. Omega Cen, 47 Tuc, M22, NGC6293. 
\end{itemize}

The results presented above suggest a new avenue for IMBH formation in GCs.  Notwithstanding, future studies aiming to reproduce these results should bear in mind the following weaknesses inherent to the MOCCA code:
\begin{itemize}
\item In the framework of the MOCCA code any dissipational processes are not taken into account. The Fewbody code (an important ingredient of the MOCCA code) models \textit{only} direct gravitational interactions between stars and binaries. The hydrodynamics, gravitational radiation (GR), tidal effects (distortions and dissipation), mass-loss, etc. are not accounted for. In general, however, we expect the inclusion of dissipational effects to increase the rate of IMBH mass buildup.
\item In the FAST scenario, we caution that the final IMBH mass is sensitive to the assumed initial maximum BH mass, and hence the maximum mass, $m_{max}$, of the stellar IMF. We assume $m_{max} = 100 M_{\odot}$ for the models presented in this paper, but have checked that IMBHs can still form via the mechanism presented here for maximum stellar IMF masses as low a few tens of $M_{\odot}$.  This caveat does not apply to the SLOW scenario, since here the initial BH mass is always small, due to forming later on in the cluster evolution from binary evolution or collisions involving NSs and/or WDs.
\item MOCCA does not cope well with physical processes that have a time scale comparable to the local dynamical time scale. Hence, for example, the formation and evolution of an IMBH loss cone can not be modelled. Having said that, MOCCA does successfully reproduce the expected IMBH signatures in the simulated VDP and SBP profiles; 
\item MOCCA was not designed to follow the dynamical evolution of extremely massive objects (larger than a few hundred $M_{\odot}$). This is due to the adopted method for computing the cluster potential. Nevertheless, the initial IMBH mass growth is modelled correctly by MOCCA;
\item MOCCA was not designed to model the relaxation of a cluster hosting a massive IMBH. This is once again connected to the adopted method of calculating the cluster potential;
\item The Fewbody code (Fregeau et al. 2004) works well for extreme mass ratios. However, doubts linger regarding Fewbody's prescription for mergers, mass accretion rates and general relativistic effects involving massive IMBHs. Again, we do not expect these issues to significantly affect our conclusions regarding the initial phase of IMBH growth; 
\item The accuracy of the BSE code \citep{HurleyPT2000,HurleyTP2002} for modelling binary evolution and mass transfer in binaries with extreme mass ratios (and/or extremely massive compact objects) is suspect. We note, however, that the mass transferred onto an IMBH because of binary/stellar evolution is not the dominant mechanism for mass growth. Thus, IMBH formation occurs even if the contribution from binary mass transfer is ignored;
\item Extremely large kicks in the case of mergers of BHs with misaligned spins. Such large kicks can spoil or weaken the FAST scenario, but not influence much the SLOW scenario.
\end{itemize}

The issues discussed above except the first one can contribute to decreasing the probability of IMBH formation, and/or lower the rate of BH mass growth, and should be re-visited in future studies using more sophisticated simulation techniques. We emphasise that the results presented in this paper are very sensitive to the assumed rates of accretion onto BHs, which are highly uncertain.  For example, our prescription for BH growth does not account for the possibility that radiation produced during the accretion process could ionise the surrounding gas and ultimately drive inverse Compton scattering.  This could substantially reduce our assumed BH accretion rates.  This issue should be re-visited when better prescriptions for BH accretion become available.  Regardless, the results presented in this paper illustrate that, for our assumed initial conditions, stellar mass BHs that remain in clusters until the present-day should be actively accreting for most of the cluster lifetime, independent of the actual accretion rates.  We have shown that BH growth could be significant, and even result in IMBH formation.

\section*{Acknowledgements}\label{sec:ack} 

The authors are grateful to John Fregeau and Jarrod Hurley for making their Fewbody and BSE codes accessible to the public and for very helpful suggestions related to efficient use of the codes. MG warmly thanks Douglas Heggie for numerous discussions and suggestions which help to better understand the process of IMBH formation and its weak sides and markedly improved the manuscript presentation. MG, AH and AA were supported by the National Science Centre through the grant DEC-2012/07/B/ST9/04412. NL gratefully acknowledges the generous support of an NSERC PDF award. The authors thank the referee Christopher Tout for a very helpful report which significantly contributed to improve the presentation of the manuscript.

\bsp

\label{lastpage}


\begin{thebibliography}{mnras}

\bibitem[\protect\citeauthoryear{Anderson \& van der Marel}{2010}]{anderson10} Anderson J., van der Marel R.~P. 2010, \apj, 710, 1032

\bibitem[\protect\citeauthoryear{Bahcall \& Wolf}{1976}]{BahcallWolf1976} Bahcall J.~N., Wolf R.~A., 1976, \apj, 209, 214

\bibitem[Banerjee \& Kroupa(2011)]{BanerjeeKroupa2011} Banerjee, S., Kroupa, P., 2011, \apjl, 741, L12

\bibitem[\protect\citeauthoryear{Baumgardt, Gualandris \& Portegies Zwart}{2006}]{BaumgardtGP2006} Baumgardt H., Gualandris A., Portegies Zwart S., 2006, \jpcs, 56, 301

\bibitem[\protect\citeauthoryear{Baumgardt, Makino \& Hut}{2005}]{BaumgardtMakinoHut2005} Baumgardt H., Makino J., Hut P., 2005, \apj, 620, 238

\bibitem[\protect\citeauthoryear{Belczynski et al.}{2014}]{Belczynskietal2014} Belczynski K., Buonanno A., Cantiello M., Fryer C.~L., Holz D.~E., Mandel I., Miller M.~C., Walczak M., 2014, \apj, 789, 120

\bibitem[\protect\citeauthoryear{Belczynski, Kalogera \&  Bulik}{2002}]{Belczynskietal2002} Belczynski K., Kalogera V., Bulik T., \apj, 572, 407

\bibitem[\protect\citeauthoryear{Bianchini et al.}{2015}]{Bianchinietal2015} Bianchini P., Norris M., van de Ven G., Schinnerer E., 2015, arXiv:1501.04114

\bibitem[\protect\citeauthoryear{Breen \& Heggie}{2013}]{BreenHeggie2013} Breen P.~G., Heggie D.~C., 2013, \mnras, 436, 584 

\bibitem[\protect\citeauthoryear{Brown, Geller \& Kenyon}{2014}]{BrownGK2014} Brown W.~R., Geller M.~J., Kenyon S.~J., 2014, \apj, 787, 89

\bibitem[\protect\citeauthoryear{Contenta, Varri \& Heggie}{2015}]{Contentaetal2015} Contenta F., Varri A.~L., Heggie D.~C., 2015, \mnras, 449, L100

\bibitem[\protect\citeauthoryear{Cseh et al.}{2010}]{Csehetal2010} Cseh D.. Kaaret P., Corbel S., K{\"o}rding E., Coriat M., Tzioumis A., Lanzoni B., 2010, \mnras, 406, 1049

\bibitem[\protect\citeauthoryear{De Marchi,  Paresce \& Pulone}{2007}]{DeMarchietal2007} De Marchi G., Paresce F., Pulone L., 2007, \apj, 656, L65

\bibitem[\protect\citeauthoryear{Ferrarese \& Merritt}{2000}]{FerrareseMerritt2000} Ferrarese L., Merritt D., 2000, \apjl, 539, L9

\bibitem[\protect\citeauthoryear{Fukushige \& Heggie}{2000}]{FukushigeHeggie2000} Fukushige T., Heggie D.~C., 2000, \mnras, 318, 753

\bibitem[\protect\citeauthoryear{Fregeau et al.}{2004}]{FCZR2004} 
Fregeau J.~M., Cheung P., Portegies Zwart S.~F., Rasio F.~A., 2004, \mnras, 352, 1

\bibitem[\protect\citeauthoryear{Fryer, Woosley \& Heger}{2001}]{FryerWH2001} Fryer C.~L., Woosley S.~E., Heger A., 2001, \apj, 550, 332

\bibitem[\protect\citeauthoryear{Gebhardt et al.}{2000}]{Gebhardtetal2000} Gebhardt K. et al., 2000, \apjl, 539, L13

\bibitem[\protect\citeauthoryear{Gebhardt, Rich \& Ho}{2005}]{gebhardt05} Gebhardt K., Rich R.~M., Ho L. ~C. 2005, \apj, 634, 1093

\bibitem[\protect\citeauthoryear{Giersz}{2001}]{Giersz2001} Giersz M., 2001, \mnras, 324, 218

\bibitem[\protect\citeauthoryear{Giersz et al.}{2013}]{Gierszetal2013} Giersz M., Heggie D.~C., Hurley J.~R., Hypki A., 2013, \mnras, 411, 2184

\bibitem[\protect\citeauthoryear{Guillochon \& Ramirez-Ruiz}{2013}]{GuillochonRamirez-Ruiz2013} Guillochon J., Ramirez-Ruiz E., 2013, \apj, 767, 25 

\bibitem[\protect\citeauthoryear{G{\"u}ltekin et al.}{2009}]{Gultekin2009} G{\"u}ltekin K., Cackett E.~M., Miller J.~M., Di Matteo T., Markoff S., Richstone D.~O., 2009, \apj, 706, 404

\bibitem[\protect\citeauthoryear{G{\"u}rkan, Freitag \& Rasio}{2004}]{Gurkan2004}
{G{\"u}rkan, M.~A., Freitag, M. \& Rasio, F.~A.} 2004, \apj, 604, 632

\bibitem[\protect\citeauthoryear{Harris}{1996}]{Harris1996} Harris W.~E., 1996, \aj, 112, 1487

\bibitem[\protect\citeauthoryear{Heggie}{2014}]{Heggie2014} Heggie, D.~C., 2014, \mnras, 445, 3435 


\bibitem[\protect\citeauthoryear{Heggie, Portegies Zwart, 
\& Hurley}{2006}]{Heggieetal2006} Heggie D.~C., Portegies Zwart S., Hurley J.~R., 2006, NewA, 12, 20 

\bibitem[\protect\citeauthoryear{H{\'e}non}{1971}]{Henon1971} H{\'e}non, M.~H., 1971, \apss, 14, 151 

\bibitem[\protect\citeauthoryear{Hurley}{2007}]{Hurley2007} Hurley J.~R., 2007, \mnras, 379, 93

\bibitem[\protect\citeauthoryear{Hurley, Pols, \& Tout}{2000}]{HurleyPT2000} Hurley J.~R., Pols O.~R., Tout C.~A., 2000, \mnras, 315, 543 

\bibitem[\protect\citeauthoryear{Hurley, Tout, \& Pols}{2002}]{HurleyTP2002} Hurley J.~R., Tout C.~A., Pols O.~R., 2002, \mnras, 329, 897 

\bibitem[\protect\citeauthoryear{Konstantinidis, Amaro-Seoane, 
\& Kokkotas}{2013}]{Konstantinidisetal2013} Konstantinidis S., Amaro-Seoane P., Kokkotas K.~D., 2013, A\&A, 557, A135 

\bibitem[\protect\citeauthoryear{Kroupa}{1995}]{Kroupa1995} Kroupa P., 1995, \mnras, 277, 1507

\bibitem[\protect\citeauthoryear{Kroupa}{2001}]{Kroupa2001} Kroupa P., 2001, \mnras, 322, 231

\bibitem[\protect\citeauthoryear{Kroupa, Tout \& Gilmore}{1993}]{Kroupaetal1993} Kroupa, P., Tout, C.~A., Gilmore, G., 1993, \mnras, 262, 545 

\bibitem[\protect\citeauthoryear{Kyutoku et al.}{2015}]{Kyutokuetal2015} Kyutoku K., Ioka K., Okawa H., Shibata M., Taniguchi K., 2015,  arXiv:1502.05402

\bibitem[\protect\citeauthoryear{Leigh et al.}{2013a}]{Leighetal2013a} 
Leigh N.~W.~C., B\"{o}ker T., Maccarone T.~J., Perets H.~B., 2013a, \mnras, 433, 1958

\bibitem[\protect\citeauthoryear{Leigh \& Geller}{2012}]{LeighGeller2012} 
Leigh N.~W.~C., Geller A.~M., 2012, \mnras, 425, 2369

\bibitem[\protect\citeauthoryear{Leigh et al.}{2013b}]{Leighetal2013b} 
Leigh N.~W.~C., Giersz T., Webb J.~J., Hypki A., De March G., Kroupa P., Sills A., 2013b, \mnras, 436, 3399

\bibitem[\protect\citeauthoryear{Leigh et al.}{2015a}]{Leighetal2015a} 
Leigh N.~W.~C., Giersz T., Marks M., Webb J.~J., Hypki A., Heinke C.~O., Kroupa P., Sills A., 2015a, \mnras, 446, 226

\bibitem[\protect\citeauthoryear{Leigh et al.}{2014b}]{Leighetal2014b} 
Leigh N.~W.~C., L\"utzgendorf N., Geller A.~M., Maccarone T.~J., Heinke C., Sesana A. 2014b, \mnras, 444, 29

\bibitem[\protect\citeauthoryear{Leigh et al.}{2014a}]{Leighetal14a} Leigh N. W. C., Mastrobuono-Battisti A., Perets H. B., B\"oker T. 2014a, \mnras, 441, 919

\bibitem[\protect\citeauthoryear{Li et al.}{2012}]{Lietal2012} Li Y., Luo A., Zhao G., Lu Y., Ren J., Zuo F., 2012 \apjl, 744, L24 

\bibitem[\protect\citeauthoryear{Long et al.}{2015}]{Longetal2015} Long W., Spurzem R., Aarseth S., Giersz M., Askar, A., Naab T., Kouwenhoven M.~B.~N., 2015 \mnras submitted 

\bibitem[\protect\citeauthoryear{Lu \& Kong}{2011}]{LuKong2011} Lu T.~N., Kong A.~ K.~H., 2011, \apj, 729, L25

\bibitem[\protect\citeauthoryear{L\"{u}tzgendorf et al.}{2013}]{Lutzgendorfetal2013} L\"{u}tzgendorf N. et al., 2013, \aap, 552, 49

\bibitem[\protect\citeauthoryear{Maccarone \& Servillat}{2008}]{maccarone08} Maccarone T.~J., Servillat M. 2008, \mnras, 389, 379

\bibitem[\protect\citeauthoryear{Madau \& Rees}{2001}]{MadauRees2001}
{Madau, P. \& Rees, M.~J.} 2001, \apjl, 551, L27

\bibitem[\protect\citeauthoryear{Magorian et al.}{1998}]{Magorianetal1998} Magorrian J. et al., 1998, \aj, 115, 2085

\bibitem[\protect\citeauthoryear{Matsumoto et a.}{2001}]{Matsumotoetal2001}  Matsumoto H. et al. 2001, \apj, 547, L25

\bibitem[\protect\citeauthoryear{Metzger \& Stone}{2015}]{MetzgerStone2015} Metzger B.~D., Stone N.~C., 2015, arXiv150603453

\bibitem[\protect\citeauthoryear{Miller-Jones et al.}{2012}]{Miller-Jonesetal2012} Miller-Jones J.~C.~A. et al., 2012, \apj, 755, L1

\bibitem[\protect\citeauthoryear{Milone et al.}{2012}]{Miloneetal2012}
Milone A.~P. et al., 2011, \aap, 540, 16

\bibitem[\protect\citeauthoryear{Noyola \& Baumgardt}{2011}]{NoyolaBaumgard2011} Noyola E., Baumgardt, H., 2011, \apj, 743, 52 

\bibitem[\protect\citeauthoryear{Noyola, Gebhardt \& Bergmann}{2008}]{noyola08} Noyola E., Gebhardt K., Bergmann M. 2008, \apj, 676, 1008

\bibitem[\protect\citeauthoryear{Palladino et al.}{2014}]{Palladinoetal2014} Palladino L.~E., Schlesinger K.~J., Holley-Bockelmann K., Allende Prieto C., Beers T.~C., Lee Y.~S., Schneider D.~P., 2014, \apj, 782, 57

\bibitem[\protect\citeauthoryear{Papish \& Perets}{2015}]{PapishPerets2015} Papish O., Perets H.~B., 2015, arXiv150203453

\bibitem[\protect\citeauthoryear{Perets \& Fabrycky}{2009}]{perets09} Perets H.~B., Fabrycky D.~C. 2009, \apj, 697, 1048

\bibitem[\protect\citeauthoryear{Portegies Zwart et al.}{2004}]{Portegiesetal2004}
Portegies Zwart, S.~F., Baumgardt, H., Hut, P., Makino, J., McMillan, S.~L.~W. 2004, \nature, 428, 724

\bibitem[\protect\citeauthoryear{Sesana, Haardt  \& Madau}{2015}]{Sesanaetal2006} Sasena A., Haardt F.., Madau P., 2006, \apj, 651, 392

\bibitem[\protect\citeauthoryear{Shiokawa et al.}{2015}]{Shiokawaetal2015} Shiokawa H., Krolik1 J.~H., Cheng R.~M., Piran T., Noble S.~C., 2015, arXiv:1501.04365

\bibitem[\protect\citeauthoryear{Soria et al.}{2010}]{Soriaetal2010} Soria R., Hau G.~K.~T., Graham A.~W., Kong A.~K.~H., Kuin N.~P.~M., Li I.~H., Liu J.-F., Wu K., 2010, \mnras, 405, 870

\bibitem[\protect\citeauthoryear{Stod\'o\l kiewicz}{1986}]{Stodolkiewicz1986} Stod\'o\l kiewicz, J.~S., 1986, Acta Astronomica, 36, 19 

\bibitem[\protect\citeauthoryear{Tauris}{2015}]{Tauris2015} Tauris T.~M., 2015, \mnras, 448, L6

\bibitem[\protect\citeauthoryear{Trenti, Vesperini \& Pasquato}{2010}]{TrentiVesperiniPasquato2010} Trenti M., Vesperini E., Pasquato M., 2010, \apj, t08, 1598

\bibitem[\protect\citeauthoryear{van der Marel \& Anderson}{2010}]{vandermarel10} van der Marel R.~P., Anderson J. 2010, \apj, 710, 1063

\bibitem[\protect\citeauthoryear{Whalen \& Fryer}{2012}]{WhalenFryer2012} Whalen D.~J., Fryer C.~L, 2012, \apj, 756, L19

\bibitem[\protect\citeauthoryear{Zheng et al.}{2014}]{Zhengetal2014} Zheng Z. et al., 2014, \apj, 785, L23

\bibitem[\protect\citeauthoryear{Zhong, Berczik \& Spurzem}{2014}]{ZhongBS2014} Zhong S.,Berczik P., Spurzem R., 2014, \apj 792, 137

\end{thebibliography}
\end{document}